\documentclass[aps,prd,groupedaddress,showpacs,eqsecnum]{revtex4}
\usepackage{graphicx}
\usepackage{dcolumn}
\usepackage{bm}

\begin{document}

\title{The decays $\omega(782)\mbox{, }\phi(1020)\to5\pi$ in the hidden local symmetry approach. }

\author{N.~N.~Achasov}
\email[]{achasov@math.nsc.ru}
\author{A.~A.~Kozhevnikov}
\email[]{kozhev@math.nsc.ru}
\altaffiliation{}
\affiliation{Laboratory of  Theoretical Physics, S.~L.~Sobolev Institute for Mathematics,
630090, Novosibirsk, Russian Federation}

\date{\today}

\begin{abstract}
The decays $\omega\to2\pi^+2\pi^-\pi^0$ and
$\omega\to\pi^+\pi^-3\pi^0$ are reconsidered in the hidden local
symmetry approach (HLS) added with the anomalous terms. The decay
amplitudes are analyzed in detail, paying the special attention to
the Adler condition of vanishing the whole amplitude at vanishing
momentum of any final pion.
Combining the Okubo-Zweig-Iizuka (OZI) rule applied to the five
pion final state, with the Adler condition, we calculate also the
$\phi\to2\pi^+2\pi^-\pi^0$ and $\phi\to\pi^+\pi^-3\pi^0$ decay
amplitudes. The partial widths of the above decays are evaluated,
and the excitation curves in $e^+e^-$ annihilation are obtained,
assuming reasonable particular relations among the parameters
characterizing the anomalous terms of the HLS Lagrangian. The
evaluated branching ratios
$B_{\phi\to\pi^+\pi^-3\pi^0}\sim2\times10^{-7}$ and
$B_{\phi\to2\pi^+2\pi^-\pi^0}\sim7\times10^{-7}$ are such that
with the luminosity $L=500\mbox{ pb}^{-1}$ attained at DA$\Phi$NE
$\phi$ factory, one may already possess about 1685 events of the
decays $\phi\to5\pi$.
\end{abstract}

\maketitle

\section{Introduction}
\label{sec1}

The purpose of the present paper is to calculate the branching
ratios of the decays
\begin{equation}
\omega\to\pi^+\pi^-3\pi^0,\label{reacn}\end{equation}
\begin{equation}
\omega\to2\pi^+2\pi^-\pi^0,\label{reacc}\end{equation}
\begin{equation}
\phi\to\pi^+\pi^-3\pi^0,\label{reacnfi}\end{equation} and
\begin{equation}
\phi\to2\pi^+2\pi^-\pi^0,\label{reaccfi}\end{equation} in the
framework of chiral model for pseudoscalar and low lying vector
mesons  based on hidden local symmetry (HLS)
\cite{hls1,hls2}. This model  incorporates vector mesons into chiral
theory in a most elegant way. The fact is that the  low energy
theorems for anomalous
processes such as, say, the decay $\pi^0\to\gamma\gamma$, are
fulfilled automatically in HLS. Since the general form of both
nonanomalous and anomalous parts of the Lagrangian is given in
Ref.~\cite{hls1,hls2},  we  write down here
the weak field limit of the above Lagrangian restricted to the
subgroup $SU(2)\times U(1)$ with the only isovector $\bm{\pi}$,
$\bm{\rho}$, and isoscalar $\omega$ mesons taken into account.
Taking into account the coupling of $\phi(1020)$ meson with the
mesons composed of nonstrange quarks demands additional
assumptions to be discussed below.

The  nonanomalous part of the HLS Lagrangian (denoted as $nan$)
is obtained from general expression of Ref.~\cite{hls1,hls2,schechter}
and looks as
\begin{eqnarray}
\cal{L}^{\rm
nan}&=&-{1\over4}\bm{\rho}_{\mu\nu}^2-{1\over4}\omega_{\mu\nu}^2+
{1\over2}ag^2f_\pi^2\left(\bm{\rho}^2_\mu+\omega^2_\mu\right)+
{1\over2}\left(\partial_\mu\bm{\pi}\right)^2-{1\over2}m^2_\pi\bm{\pi}^2+
{m^2_\pi\over24f^2_\pi}\bm{\pi}^4+   \nonumber\\
&&{1\over2f^2_\pi}\left({a\over4}-{1\over3}\right)
[\bm{\pi}\times\partial_\mu \bm{\pi}]^2+
{1\over2}ag\left(1-{\bm{\pi}^2\over12f^2_\pi}\right)\left(\bm{\rho}_\mu\cdot
[\bm{\pi}\times\partial_\mu\bm{\pi}]\right), \label{lnan}
\end{eqnarray}
where the dot ($\cdot$) and cross ($\times$) stand, respectively,
for the scalar and vector products in the isotopic space,
\begin{eqnarray}
\bm{\rho}_{\mu\nu}&=&\partial_\mu\bm{\rho}_\nu-\partial_\nu\bm{\rho}_\mu+
g[\bm{\rho}_\mu\times\bm{\rho}_\nu],     \nonumber\\
\omega_{\mu\nu}&=&\partial_\mu\omega_\nu-\partial_\nu\omega_\mu
\label{strength}
\end{eqnarray}
are, respectively,  the field strengths of the isovector field
$\bm{\rho}_\mu$ and the isoscalar field $\omega_\mu$, $g$ is the
gauge coupling constant, $f_\pi=92.4$ MeV is the pion decay
constant, while $a$ is HLS parameter. The boldface characters
refer hereafter to the vectors in isotopic space. As is clear from
Eq.~(\ref{lnan}),
\begin{eqnarray}
g_{\rho\pi\pi}&=&{1\over2}ag,   \nonumber\\
m_\rho^2&=&ag^2f_\pi^2\label{rhoparam}\end{eqnarray} are the
$\rho\pi\pi$ coupling constant and the $\rho$ mass
squared, respectively. The $\omega(782)$ is degenerate with $\rho$
in the present model. Note that  $a=2$, if one requires the
universality condition $g=g_{\rho\pi\pi}$. Then
the Kawarabayashi-Suzuki-Riazzuddin- Fayyazuddin (KSRF)
relation \cite{ksrf} arises
\begin{equation}
{2g^2_{\rho\pi\pi}f^2_\pi\over m^2_\rho}=1 \label{ksrf}
\end{equation}
which beautifully agrees with experiment. The $\rho\pi\pi$
coupling constant resulting from this relation is
$g_{\rho\pi\pi}=5.9$.

To include the decays of the $\omega$ meson to the many pion
states one should add the anomalous terms (denoted below as $an$).
They are given in Ref.~\cite{hls1,hls2}. Since only strong decays
will be of our concern here, we omit the terms containing
electromagnetic field. Again, restricting ourselves by the weak
field limit and by the $\rho$, $\omega$, and $\pi$ fields, we
arrive at the expression
\begin{eqnarray}
\cal{L}^{\rm
an}&=&{n_cg\over32\pi^2f^3_\pi}(c_1-c_2-c_3)\varepsilon_{\mu\nu\lambda\sigma}\omega_\mu
\left(\partial_\nu\bm{\pi}\cdot[\partial_\lambda\bm{\pi}\times\partial_\sigma\bm{\pi}]
\right)+        \nonumber\\ &&
{n_cg\over128\pi^2f^5_\pi}\left[-c_1+{5\over3}\left(c_2+c_3\right)\right]
\varepsilon_{\mu\nu\lambda\sigma}\omega_\mu
\left(\partial_\nu\bm{\pi}\cdot[\partial_\lambda\bm{\pi}\times\partial_\sigma\bm{\pi}]
\right)\bm{\pi}^2-        \nonumber\\ &&
{n_cg^2c_3\over8\pi^2f_\pi}\varepsilon_{\mu\nu\lambda\sigma}\partial_\mu\omega_\nu
\left\{\left(\bm{\rho}_\lambda\cdot\partial_\sigma\bm{\pi}\right)+{1\over6f^2_\pi}
\left[\left(\bm{\rho}_\lambda\cdot\bm{\pi}\right)\left(\bm{\pi}\cdot\partial_\sigma
\bm{\pi}\right)-\bm{\pi}^2\left(\bm{\rho}_\lambda\cdot\partial_\sigma\bm{\pi}\right)
\right]\right\}-     \nonumber\\ &&
{n_cg^2\over8\pi^2f_\pi}(c_1+c_2-c_3)\varepsilon_{\mu\nu\lambda\sigma}\omega_\mu
\left\{\frac{1}{4f^2_\pi}\left(\partial_\nu\bm{\pi}\cdot\bm{\rho}_\lambda\right)\left(\bm{\pi}\cdot
\partial_\sigma\bm{\pi}\right)-{g\over4}\left(\left[\bm{\rho}_\nu\times
\bm{\rho}_\lambda\right]\cdot\partial_\sigma\bm{\pi}\right)\right\},
\label{lan}
\end{eqnarray}
where $n_c=3$ is the number of colors, $c_{1,2,3}$ are arbitrary
constants multiplying three independent structures in the solution
\cite{hls1,hls2} of the Wess-Zumino anomaly equation \cite{wza};
the fourth constant $c_4$ multiplying the structure that includes
electromagnetic field, as is explained above, is dropped. Our
normalization of $c_{1,2,3}$ is in accord with Ref.~\cite{hls2}.
As is evident from the third line of Eq.~(\ref{lan}), the
$\omega\rho\pi$ coupling constant is
\begin{equation}
g_{\omega\rho\pi}=-{n_cg^2c_3\over8\pi^2f_\pi}.
\label{gomrp}\end{equation} Assuming
\begin{equation}
c_1-c_2-c_3=0,\label{rel1}\end{equation} i.e. the absence of the
point like $\omega\to\pi^+\pi^-\pi^0$ amplitude, and using the
$\omega\to\pi^+\pi^-\pi^0$ partial width to extract
$g_{\omega\rho\pi}$, the $\rho\to\pi^+\pi^-$ partial width and
Eq.~ (\ref{rhoparam}) to extract $g=g_{\rho\pi\pi}=6.00\pm0.01$
(assuming $a=2$), one finds
\begin{equation}
c_3=0.99\pm0.01,\label{c3}\end{equation} where the errors come
from the errors of the $\omega$ and $\rho$ widths. Hereafter we
use the particle parameters (masses, full and partial widths etc.)
taken from Ref.~\cite{pdg}.

The other material of the paper is organized as follows. Section
\ref{omegaamp} is devoted to obtaining the
$\omega\to\pi^+\pi^-3\pi^0$ and $\omega\to2\pi^+2\pi^-\pi^0$ decay
amplitudes from the Lagrangians given by Eq.~(\ref{lnan}) and
(\ref{lan}) and verifying the Adler condition for their
expressions. The results of the evaluation of the branching ratios
at the $\omega$ pole position and the calculation of the
excitation curves of the above decays in $e^+e^-$ annihilation are
given in Sec.~\ref{omegaresults}, imposing the natural
constraints on the parameters $c_{1,2,3}$ characterizing the anomalous
terms of the HLS Lagrangian.
As is shown there, the evaluated branching ratios  depend
insignificantly on the exact form of the constraints.
The reason of disagreement with our previous evaluations
\cite{ach00} of the branching ratios for the decays (\ref{reacn})
and (\ref{reacc}) is explained. In Sec.~\ref{phiresults}, guided
by the specific assumptions about how the OZI rule is violated in
the decays of $\phi$ meson into the states containing no particles
with strangeness, the effective Lagrangian for the
$\phi\to\pi^+\pi^-3\pi^0$ and $\phi\to2\pi^+2\pi^-\pi^0$ decay
amplitudes is written. Under the assumptions about the free
parameters of this Lagrangian similar to $c_{1,2,3}$, the
branching ratios and the $e^+e^-$ annihilation excitation curves
for the five pion decays of the $\phi$ are given in the same
Section. The estimates of the number of events of the decays
$\omega,\phi\to\pi^+\pi^-3\pi^0$ and
$\omega,\phi\to2\pi^+2\pi^-\pi^0$ at the respective $\omega$ and
$\phi$ peak positions and the general conclusions about the
possibilities of detecting the decays under consideration in
$e^+e^-$ annihilation are given in Sec.~\ref{conclu}. Kinematical
relations expressing the Lorentz  scalar products of the pion
momenta through invariant Mandelstam-like variables which are
necessary for the phase space integration, are given in Appendix.

\section{The $\omega\to\pi^+\pi^-3\pi^0$ and
$\omega\to2\pi^+2\pi^-\pi^0$ decay amplitudes} \label{omegaamp}

In this section, we obtain the $\omega\to\pi^+\pi^-3\pi^0$ and
$\omega\to2\pi^+2\pi^-\pi^0$ decay amplitudes and study their
Adler limit, i.e. the limit at the  vanishing  four-momentum of
any final pion.
Our notation for the Lorentz scalar product of two different
four-vectors $a$ and $b$ is $(a,b)=a_0b_0-({\bf a}\cdot{\bf b})$,
while the Lorentz square is denoted as usual $a^2$. We divide the
presentation into two subsections for each above mentioned
isotopic configuration of the final state pions.

\subsection{The $\omega\to\pi^+\pi^-3\pi^0$ final state}
\label{subsecomn}

\begin{figure}
\includegraphics{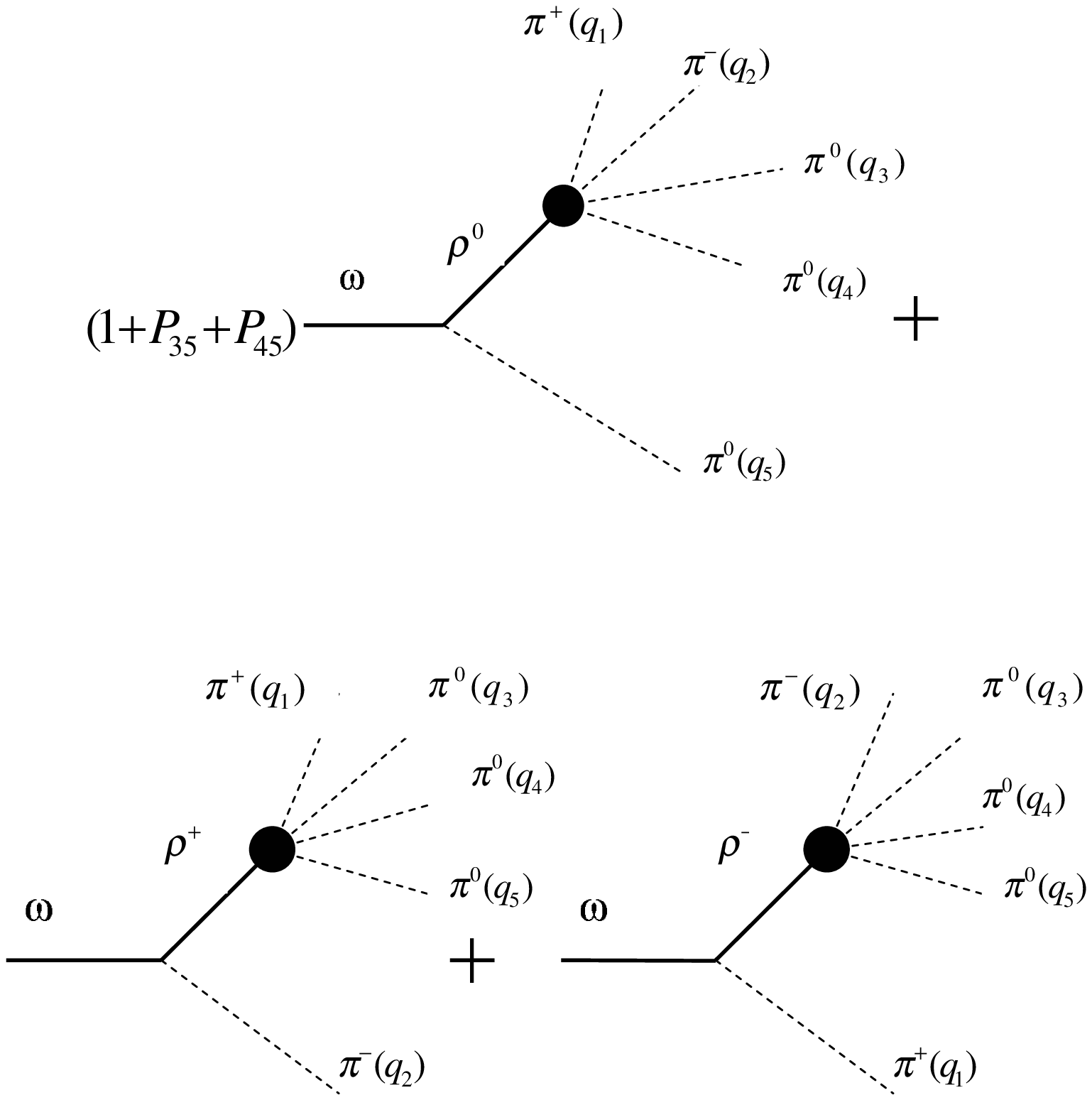}
\caption{\label{fig1}The diagrams describing the amplitudes of the decay
$\omega\to\pi^+\pi^-3\pi^0$ through the $\rho$ intermediate state
followed by the decay $\rho\to4\pi$. The shaded circles refer to
the whole $\rho\to4\pi$ amplitudes.}
\end{figure}

\begin{figure}
\includegraphics{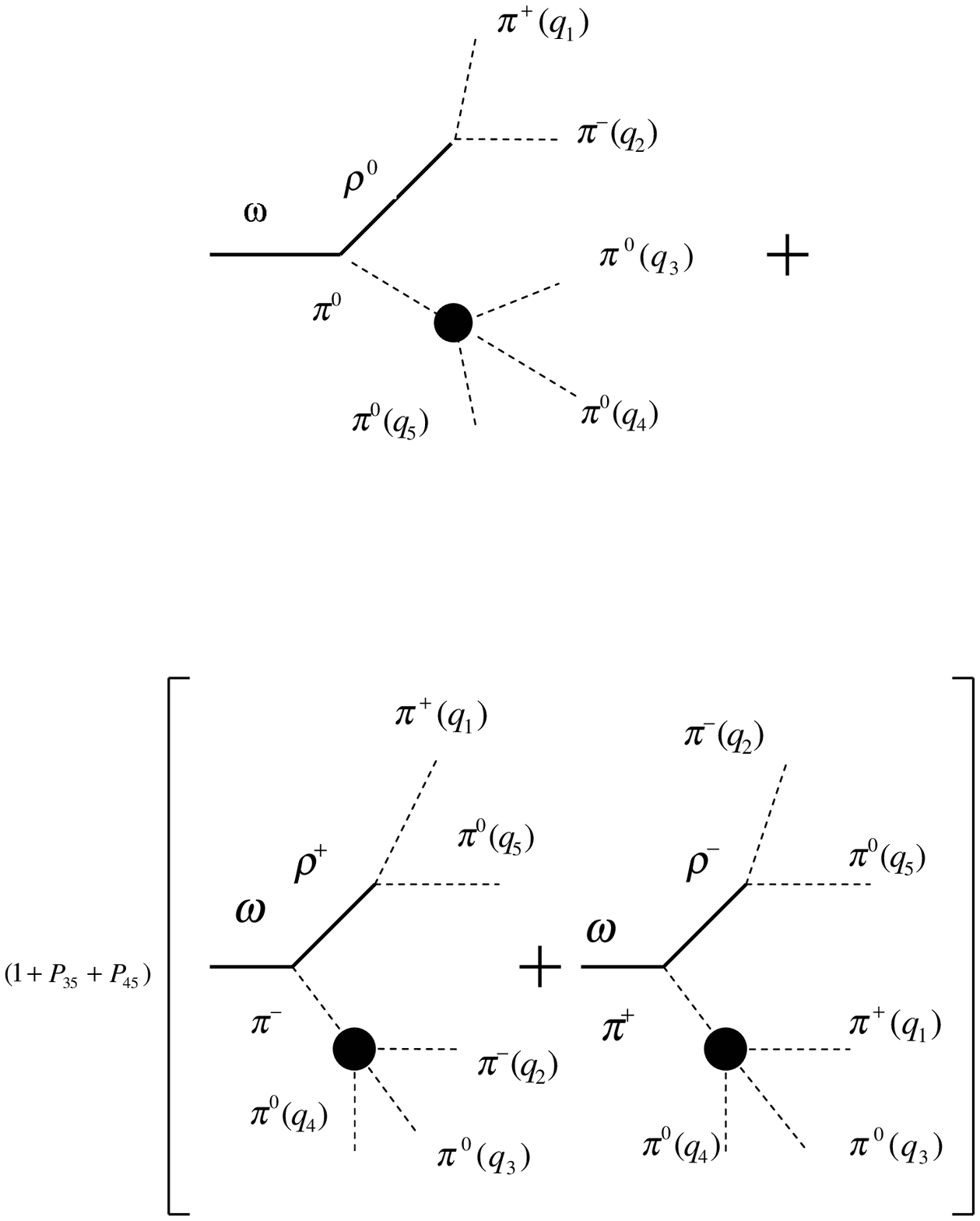}
\caption{\label{fig2}The diagrams describing the amplitudes of the decay
$\omega\to\pi^+\pi^-3\pi^0$ through the $\rho\pi$ intermediate
state followed by the transitions $\rho\to2\pi$ and $\pi\to3\pi$.
The shaded circles refer to the effective $\pi\to3\pi$ vertices
given by Eq.~(\protect\ref{4pi}). Note that non-$\pi$-pole term is
included to the  diagrams in Fig.~\protect\ref{fig4} below.}
\end{figure}

\begin{figure}
\includegraphics{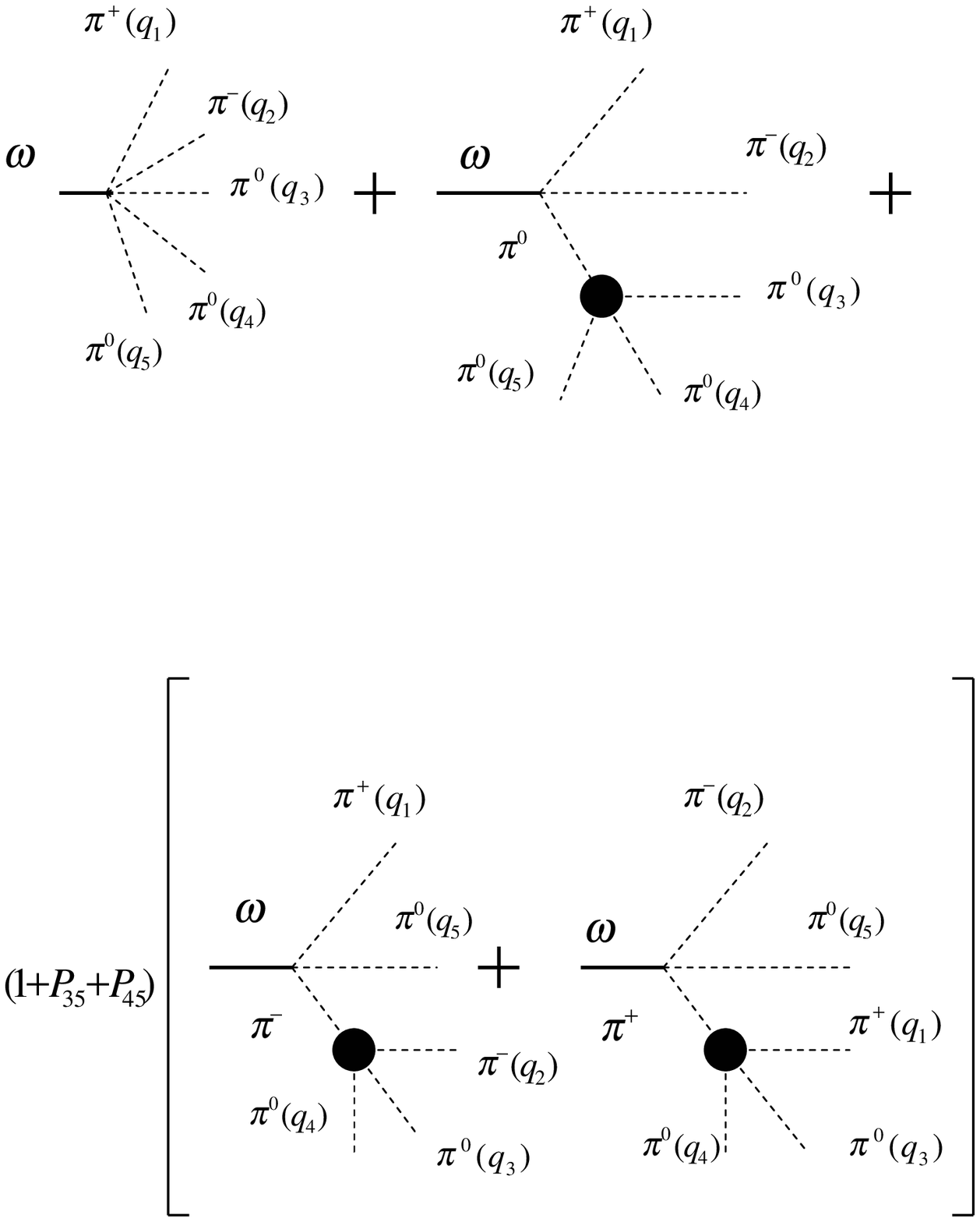}
\caption{\label{fig3}The
diagrams describing the contributions to the
$\omega\to\pi^+\pi^-3\pi^0$ decay amplitude via point like
vertices. The shaded circles refer to the effective $\pi\to3\pi$
vertices given by Eq.~(\protect\ref{4pi}).}
\end{figure}

\begin{figure}
\includegraphics{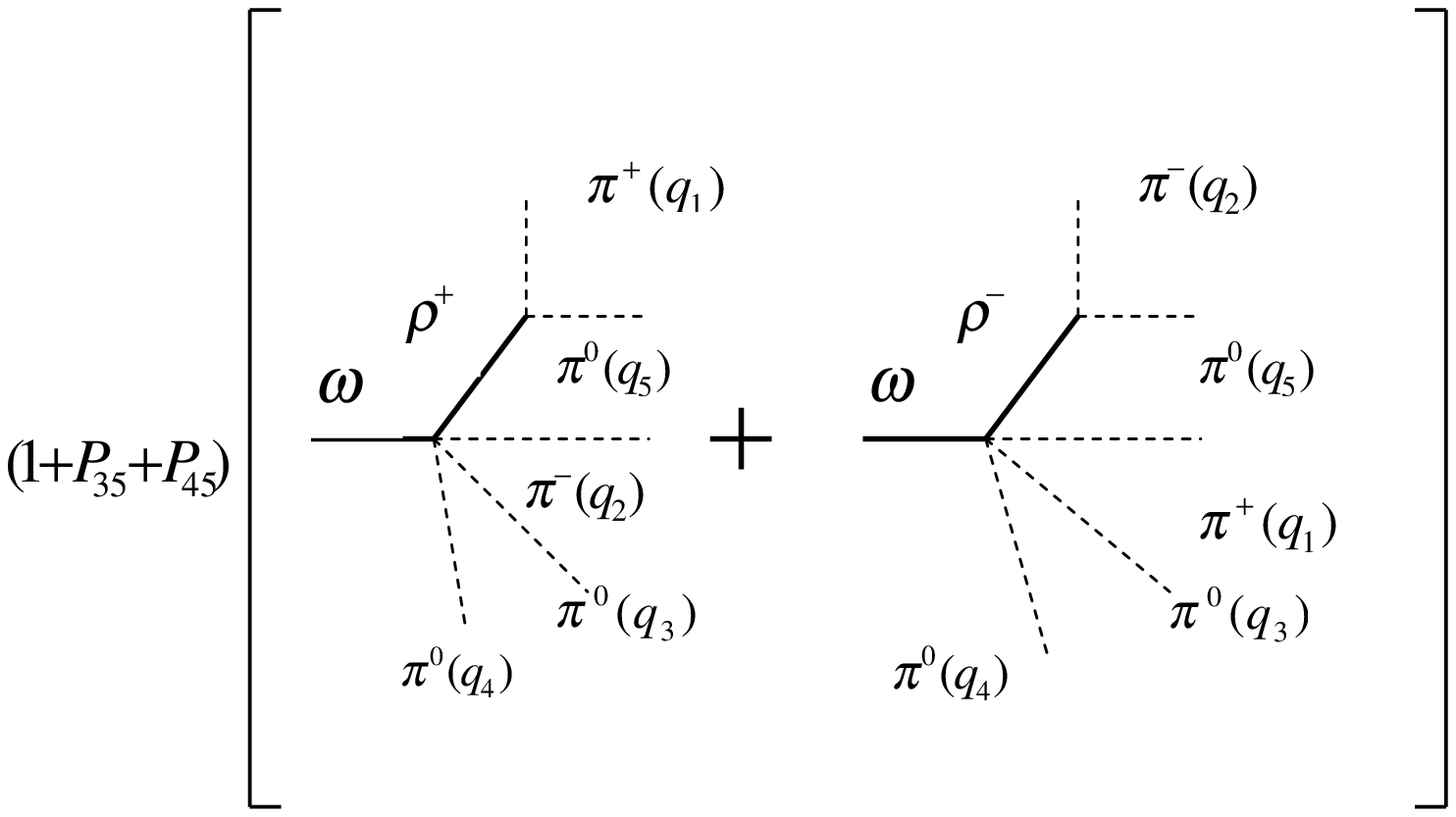}
\caption{\label{fig4}The
contributions to the $\omega\to\pi^+\pi^-3\pi^0$ decay amplitude
arising due to the chiral vertex $\omega\to\rho 3\pi$.}
\end{figure}

\begin{figure}
\includegraphics{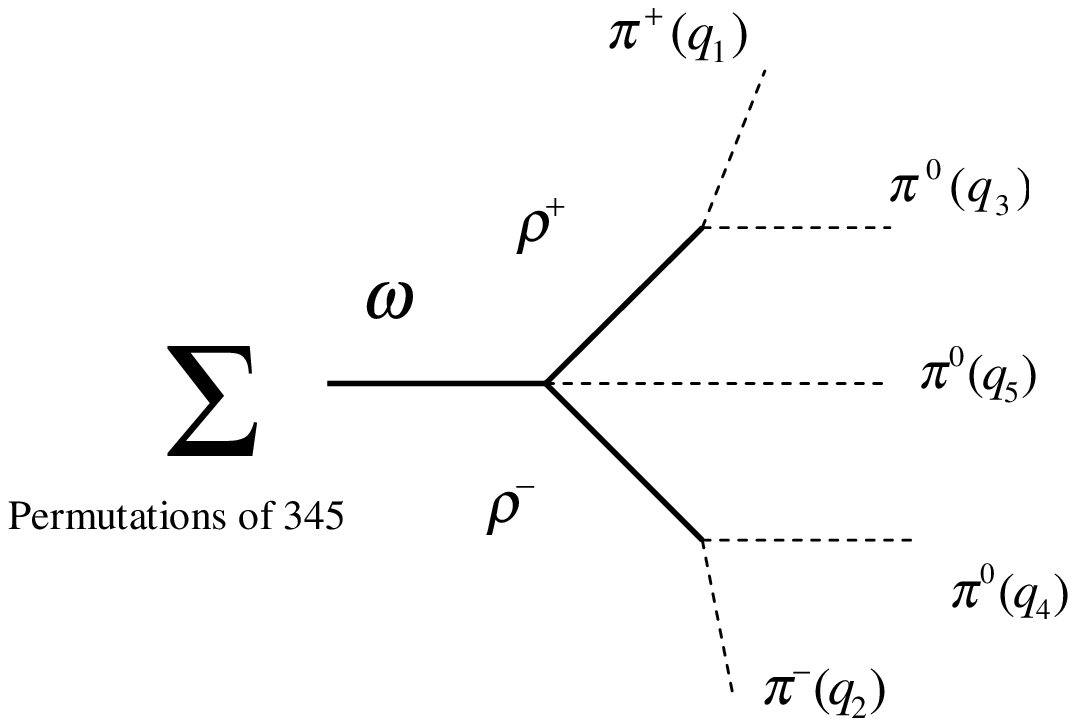}
\caption{\label{fig5}The
contributions to the $\omega\to\pi^+\pi^-3\pi^0$ decay amplitude
via intermediate state with two $\rho$ mesons. Total number of
diagrams of this kind is $3!=6$.}
\end{figure}

The diagrams for the amplitude of the decay \begin{equation}
\omega_q\to\pi^+_{q_1}\pi^-_{q_2}\pi^0_{q_3}\pi^0_{q_4}\pi^0_{q_5},
\label{omegan}\end{equation} where we explicitly label each
particle in the reaction by its four-momentum, are shown in
Fig.~\ref{fig1}$-$\ref{fig4}. Let us give the expressions
corresponding to them. The upper index $(n)$ (pointing to {\it
neutral}, because three neutral pions are in the final state) will
designate this particular isotopic state. The amplitude
corresponding to
Fig.~\ref{fig1}, includes the four pion decay of the intermediate
$\rho$ meson which was extensively discussed in, e.g., \cite{ach00}.
The Lagrangian due to Weinberg \cite{weinberg68} was used
in Ref.~\cite{ach00} to find the expressions for the $\rho\to4\pi$
decay amplitudes. This Lagrangian is different in coefficients as
compared to Eq.~(\ref{lnan}) above.  However, one can show by direct
computation that due to the  well known parameter
independence the $\rho\to4\pi$ decay amplitudes resulting from the
above Lagrangians coincide. The reason is that the terms $\propto
D_\pi(k)$ in the $\pi\to3\pi$ amplitudes,
\begin{eqnarray}
M(\pi_k^+\to\pi^+_{q_1}\pi^+_{q_2}\pi^-_{q_3})&=&{1\over2f^2_\pi}(1+\hat{P}_{12})
\left[-a(q_1,q_3)+(a-2)(q_1,q_2)+am^2_\rho{(q_2,q_3-q_1)\over
D_\rho(q_1+q_3)}-\right.\nonumber\\
&&\left.{1\over3}D_{\pi^+}(k)\right], \nonumber\\
M(\pi_k^+\to\pi^+_{q_1}\pi^0_{q_3}\pi^0_{q_4})&=&{1\over2f^2_\pi}(1+\hat{P}_{34})
\left[-(a-1)(q_3,q_4)+(a-2)(q_1,q_3)+am^2_\rho{(q_4,q_3-q_1)\over
D_\rho(q_1+q_3)}-\right. \nonumber\\
&&\left.{1\over6}D_{\pi^+}(k)\right],  \nonumber\\
M(\pi_k^0\to\pi^+_{q_1}\pi^-_{q_2}\pi^0_{q_5})&=&{1\over2f^2_\pi}(1+\hat{P}_{12})
\left[-(a-1)(q_1,q_2)+(a-2)(q_1,q_5)+am^2_\rho{(q_2,q_1-q_5)\over
D_\rho(q_1+q_5)}-\right. \nonumber\\
&&\left.{1\over6}D_{\pi^0}(k)\right], \nonumber\\
M(\pi^0\to\pi^0_{q_3}\pi^0_{q_4}\pi^0_{q_5})&=&{m^2_{\pi^0}\over
f^2_\pi}, \label{4pi}\end{eqnarray} which vanish on the pion mass
shell, give the non-$\pi$-pole terms in the $\rho\to2\pi\to4\pi$
amplitude. When  added to the point like $\rho\to4\pi$ amplitude,
they make their sum parameter independent. The same occurs with
such terms in the expression derived from Fig.~\ref{fig2} below,
which should be added to the expression derived from
Fig.~\ref{fig4}. The final expressions for the full
$\omega\to\pi^+\pi^-3\pi^0$ decay amplitude will be given below.
Hereafter $\hat{P}_{ij}$ is the operator of permutation of the
pion momenta $q_i$ and $q_j$,
\begin{eqnarray}
&&D_\rho(k)=m^2_\rho-k^2-i\sqrt{k^2}\Gamma_{\rho\to\pi^+\pi^-}(k^2),
\nonumber\\
&&\Gamma_{\rho\to\pi^+\pi^-}(k^2)={g^2_{\rho\pi\pi}\over48\pi k^2}
\left(k^2-4m^2_{\pi^+}\right)^{3/2} \label{propro}\end{eqnarray}
are the inverse propagator of $\rho$ meson and its two pion decay
width, respectively, and
\begin{equation}
D_{\pi^{+,0}}(k)=m^2_{\pi^{+,0}}-k^2 \label{proppi}\end{equation}
is the inverse propagator of $\pi^{\pm,0}$ meson.  The following shorthand
notations for inverse propagators of the particle $A$ will be
used:
\begin{equation}
D_{Aab}\equiv D_A(q_a+q_b)\mbox{, }D_{Aabc}\equiv
D_A(q_a+q_b+q_c). \label{shorthand}
\end{equation}
Let us give the expression for each diagram in
Fig.~\ref{fig1}$-$\ref{fig5}. Choosing $q_\mu\mbox{,
}\epsilon_\mu$ for the four-momentum and four-vector of
polarization of the $\omega$, one obtains
\begin{eqnarray}
M^{(n)}_1&=&{g_{\omega\rho\pi}g_{\rho\pi\pi}\over
f^2_\pi}\varepsilon_{\mu\nu\lambda\sigma}q_\mu\epsilon_\nu
\left[\left(1+\hat{P}_{35}+\hat{P}_{45}\right){q_{5\lambda}\over
D_\rho(q-q_5)}J_\sigma(\rho^0\to\pi^+_{q_1}\pi^-_{q_2}\pi^0_{q_3}\pi^0_{q_4})+
\right.\nonumber\\ &&\left.(1-\hat{P}_{12}){q_{2\lambda}\over
D_\rho(q-q_2)}J_\sigma(\rho^+\to\pi^+_{q_1}\pi^0_{q_3}\pi^0_{q_4}\pi^0_{q_5})\right]
\label{m1n}
\end{eqnarray}
for the diagram Fig.~\ref{fig1}. The
$\rho\to4\pi$ decay currents standing in Eq.~(\ref{m1n}) are
\cite{ach00}
\begin{eqnarray}
J_\sigma(\rho^0\to\pi^+_{q_1}\pi^-_{q_2}\pi^0_{q_3}\pi^0_{q_4})&=&(1-\hat{P}_{12})
(1+\hat{P}_{34})\left\{q_{1\sigma}\left(-{1\over4}+{1\over
D_{\pi^+234}}\left[(q_3,q_4-2q_2)+a\times\right.\right.\right.\nonumber\\
&&\left.\left.\left.(q_3,q_2-q_4)\left({m^2_\rho\over
D_{\rho24}}-1\right)\right]\right)+
{m^2_\rho\over2D_{\rho13}D_{\rho24}}\left[(q_3+q_1)_\sigma\times
\right.\right.\nonumber\\
&&\left.\left.(q_1-q_3,q_2-q_4)+2(q_3-q_1)_\sigma(q_1+q_3,q_2-q_4)\right]+
\right.\nonumber\\
&&\left.2\left({n_cg^2c_3\over8\pi^2}\right)^2{1\over
D_{\omega123}}\left({1\over D_{\rho12}}+{1\over
D_{\rho13}}+{1\over D_{\rho23}}+3{c_1-c_2-c_3\over
2c_3m^2_\rho}\right)\times\right.\nonumber\\
&&\left.\left[q_{1\sigma}\left((k,q_2)(q_3,q_4)-(k,q_3)(q_2,q_4)\right)+
q_{3\sigma}(k,q_1)(q_2,q_4)\right]\right\}, \label{j0}
\end{eqnarray}
(with $k=q_1+q_2+q_3+q_4$), where \begin{equation} D_{\omega
abc}\equiv
D_\omega(q_a+q_b+q_c)=m^2_\omega-(q_a+q_b+q_c)^2-im_\omega\Gamma_\omega
\label{propom}\end{equation} is the inverse propagator of the
$\omega$ (we take the fixed width approximation for $\omega$ meson
because the  $\omega$ resonance is narrow), and
\begin{eqnarray}
J_\sigma(\rho^+\to\pi^+_{q_1}\pi^0_{q_3}\pi^0_{q_4}\pi^0_{q_5})&=&(1+\hat{P}_{34}+\hat{P}_{35})
\left\{{1\over3}q_{1\sigma}\left(1-{2m^2_{\pi^0}\over
D_{\pi^0345}}\right)+{q_{3\sigma}\over
D_{\pi^+145}}(1+\hat{P}_{45})\times\right.\nonumber\\
&&\left.\left[(q_4,q_5-2q_1)+a(q_4,q_5-q_1)\left({m^2_\rho\over
D_{\rho15}}-1\right)\right]\right\}. \label{jpl}
\end{eqnarray}
The expression for the diagrams in Fig.~\ref{fig2} is
\begin{eqnarray}
M^{(n)}_2&=&-{g_{\omega\rho\pi}g_{\rho\pi\pi}\over
f^2_\pi}(1-\hat{P}_{12})(1+\hat{P}_{35}+\hat{P}_{45})(1+\hat{P}_{34})
\varepsilon_{\mu\nu\lambda\sigma}q_\mu\epsilon_\nu\times\nonumber\\
&&\left\{{q_{1\lambda}q_{5\sigma}\over
D_{\rho15}D_{\pi^+234}}\left[(q_3,q_4-2q_2)+a(q_3,q_2-q_4)\left({m^2_\rho\over
D_{\rho24}}-1\right)\right]-{q_{1\lambda}q_{2\sigma}m^2_{\pi^0}\over6
D_{\rho12}D_{\pi^0345}}\right\}. \label{m2n}
\end{eqnarray}
The expression for the diagram Fig.~\ref{fig3} is
\begin{eqnarray}
M^{(n)}_3&=&{n_cg\over32\pi^2f^5_\pi}(1-\hat{P}_{12})(1+\hat{P}_{35}+\hat{P}_{45})\varepsilon
_{\mu\nu\lambda\sigma}q_\mu\epsilon_\nu\left\{{4c_1-5(c_2+c_3)\over3}q_{1\lambda}q_{2\sigma}
+\right.\nonumber\\
&&\left.3(c_1-c_2-c_3)(1+\hat{P}_{34})\left[{q_{1\lambda}q_{5\sigma}\over
D_{\pi^+234}}\left((q_3,q_4-2q_2)+a(q_3,q_2-q_4)\left({m^2_\rho\over
D_{\rho24}}-1\right)\right)+\right.\right. \nonumber\\
&&\left.\left.{q_{1\lambda}q_{2\sigma}\over3D_{\pi^0345}}(q_3,q_4)\right]\right\}.
\label{m3n}
\end{eqnarray}
Notice the relation
\begin{equation}{n_cg\over32\pi^2f^5_\pi}=-{g_{\omega\rho\pi}g_{\rho\pi\pi}\over f^2_\pi}
\cdot{1\over2c_3m^2_\rho} \label{relcomp} \end{equation} which is
useful for an easier comparison of the present contribution with
others. The expression for the diagrams Fig.~\ref{fig4} and
\ref{fig5} are, respectively,
\begin{eqnarray}
M^{(n)}_4&=&-{g_{\omega\rho\pi}g_{\rho\pi\pi}\over
f^2_\pi}(1-\hat{P}_{12})(1+\hat{P}_{35}+\hat{P}_{45})\varepsilon
_{\mu\nu\lambda\sigma}{\epsilon_\nu(q_1-q_5)_\lambda\over2D_{\rho15}}\times\nonumber\\
&& \left[q_\mu q_{2\sigma}-{c_1+c_2-c_3\over2
c_3}q_{2\mu}(q_3+q_4)_\sigma\right], \label{m4n}\end{eqnarray} and
\begin{equation}
M^{(n)}_5={g_{\omega\rho\pi}g_{\rho\pi\pi}m^2_\rho\over
f^2_\pi}\cdot
{c_1+c_2-c_3\over4c_3}(1-\hat{P}_{12})(1+\hat{P}_{35}+\hat{P}_{45})\varepsilon
_{\mu\nu\lambda\sigma}{\epsilon_\nu(q_1-q_3)_\mu(q_2-q_4)_\lambda
q_{5\sigma}\over D_{\rho13}D_{\rho24}}.\label{m5n}\end{equation}
The full $\omega\to\pi^+\pi^-3\pi^0$ decay amplitude is
\begin{equation}M(\omega_q\to\pi^+_{q_1}\pi^-_{q_2}\pi^0_{q_3}\pi^0_{q_4}\pi^0_{q_5})=
M^{(n)}_1+M^{(n)}_2+M^{(n)}_3+M^{(n)}_4+M^{(n)}_5.
\label{mfulln}\end{equation} Since  the expression for the
amplitude is very cumbersome,  one should invoke the method
of the control of the calculations. We take the Adler condition as the
method of such a control.

\subsubsection{Verifying the Adler condition for
the $\omega\to\pi^+\pi^-3\pi^0$ decay amplitude} \label{subsubn}

The Adler condition is the condition of vanishing of the amplitude
of the process with soft pions when the four-momentum of any pion is
vanishing. Pions emitted in the decay $\omega\to5\pi$ \cite{ach00}
are truly soft, because they possess the typical momentum $|{\bf
q}_\pi|\simeq0.5m_\pi$. To verify the Adler condition, we
set any particular pion momentum to zero.  The correct
expression for the amplitude should then vanish.

i) $q_1=0$. The contributions of the diagrams Fig.~\ref{fig3},
\ref{fig4}, \ref{fig5} vanish, the contributions of the diagrams
Fig.~\ref{fig1} and \ref{fig2} are equal in magnitude but opposite
in sign, hence they are cancelled. The Adler condition is
fulfilled. The case $q_2=0$ is obtained from the case of $q_1=0$
by the permutation property, see the operator $1-\hat{P}_{12}$ in
front of each expression in Eq.~(\ref{m1n}), (\ref{j0}),
(\ref{m2n}), (\ref{m3n}), (\ref{m4n}) and (\ref{m5n}).

ii) $q_3=0$. Here the situation is more subtle. Let us represent
the amplitude at $q_3=0$ in the form $$
M^{(n)}(q_3=0)=-{g_{\omega\rho\pi}g_{\rho\pi\pi}\over
f^2_\pi}(1-\hat{P}_{12})(1+\hat{P}_{45})\epsilon_{\mu\nu\lambda\sigma}\epsilon_\nu
T_{\mu\lambda\sigma}. $$ Then one obtains the following
contributions to the tensor $T_{\mu\lambda\sigma}$ from the
diagrams Fig.~\ref{fig1}$-$\ref{fig5}, respectively:
\begin{eqnarray}
T_{\mu\lambda\sigma}^{(1)}&=&{q_\mu(q_1-q_4)_\lambda
q_{5\sigma}\over2D_{\rho14}},           \nonumber\\
T_{\mu\lambda\sigma}^{(2)}&=&{q_\mu q_{1\lambda}q_{4\sigma}\over
D_{\rho14}},\nonumber\\
T_{\mu\lambda\sigma}^{(3)}&=&-{1\over4m^2_\rho}\left(q_\mu q_{1\lambda}q_{2\sigma}+
{c_1+c_2-c_3\over c_3}\cdot q_{1\mu}q_{2\lambda}q_{4\sigma}\right),\nonumber\\
T_{\mu\lambda\sigma}^{(4)}&=&{1\over6}q_\mu
\left[{3q_{1\lambda}q_{2\sigma}\over2m^2_\rho}+{(q_1-q_4)_\lambda(2q_2-q_5)_\sigma
-2q_{1\lambda}q_{4\sigma}\over D_{\rho14}}\right]-\nonumber\\
&&{c_1+c_2-c_3\over4c_3}\left[{q_{2\mu}(q_1-q_4)_\lambda
q_{5\sigma}\over D_{\rho14}}-{q_{1\mu}q_{2\lambda}q_{4\sigma}\over
m^2_\rho}\right],      \nonumber\\
T_{\mu\lambda\sigma}^{(5)}&=&{c_1+c_2-c_3\over4c_3}\cdot{q_{2\mu}
(q_1-q_4)_\lambda q_{5\sigma}\over D_{\rho14}}. \label{adlern}
\end{eqnarray} In the above formulas, the
upper index points to the label of the corresponding figure.
Note that when obtaining the contribution
$T^{(3)}_{\mu\lambda\sigma}$, the relation Eq.~(\ref{relcomp}) is
essential. As is seen from Eq.~(\ref{adlern}), the terms with
the factor $(c_1+c_2-c_3)$ and without such a factor are cancelled
separately in the sum. Let us check this for the terms $\propto
q_\mu/6D_{\rho14}$. One has for the sum of these terms
$2(q_1-q_4)_\lambda(q_2+q_5)_\sigma+4q_{1\lambda}q_{4\sigma}$.
Using the four-momentum conservation and taking into account the
tensor $\epsilon_{\mu\nu\lambda\sigma}$, one can see that the
above momentum combination vanishes.
Hence, the Adler condition is satisfied in
the case $q_3=0$, too. The cases $q_{4,5}=0$ are obtained from
this case by Bose symmetry.

\subsection{The $\omega\to2\pi^+2\pi^-\pi^0$ final state}
\label{subsecomc}

The diagrams for the amplitude of the decay \begin{equation}
\omega_q\to\pi^+_{q_1}\pi^+_{q_2}\pi^-_{q_3}\pi^-_{q_4}\pi^0_{q_5},
\label{omegac}\end{equation} where the particles are labeled by
their four-momenta, are shown in Fig.~\ref{fig6}$-$\ref{fig10}.
Let us give the expressions corresponding to them. The upper index
$(c)$ (denoting  {\it charged}, because most pions in final
state are charged) will designate this particular isotopic state.
The expression for the diagram Fig.~\ref{fig6} looks as
\begin{eqnarray}
M^{(c)}_1&=&{g_{\omega\rho\pi}g_{\rho\pi\pi}\over
f^2_\pi}\varepsilon_{\mu\nu\lambda\sigma}q_\mu\epsilon_\nu
\left[{q_{5\lambda}\over
D_\rho(q-q_5)}J_\sigma(\rho^0\to\pi^+_{q_1}\pi^+_{q_2}\pi^-_{q_3}\pi^-_{q_4})+
\right.\nonumber\\ &&\left.(1+\hat{P}_{34}){q_{4\lambda}\over
D_\rho(q-q_4)}J_\sigma(\rho^+\to\pi^+_{q_1}\pi^+_{q_2}\pi^-_{q_3}\pi^0_{q_5})
+\right.\nonumber\\&&\left.(1+\hat{P}_{12}){q_{2\lambda}\over
D_\rho(q-q_2)}J_\sigma(\rho^-\to\pi^+_{q_1}\pi^-_{q_3}\pi^-_{q_4}\pi^0_{q_5})\right].
\label{m1c}
\end{eqnarray}
Here the currents responsible for the four pion decay of
intermediate $\rho$ meson are the following \cite{ach00}:
\begin{eqnarray}
J_\sigma(\rho^0\to\pi^+_{q_1}\pi^+_{q_2}\pi^-_{q_3}\pi^-_{q_4})&=&
(1+\hat{P}_{12})(1+\hat{P}_{34})(1-\hat{P}_{13}\hat{P}_{24})
\left\{q_{1\sigma}\left(-{1\over2}+{1\over
D_{\pi^+234}}\times\right.\right.\nonumber\\
&&\left.\left.\left[a(q_3,q_2-q_4)\left({m^2_\rho\over
D_{\rho24}}-1\right)-2(q_3,q_4)\right]\right)\right\};
\label{j0c}\end{eqnarray} and
\begin{eqnarray}
J_\sigma(\rho^+\to\pi^+_{q_1}\pi^+_{q_2}\pi^-_{q_3}\pi^0_{q_5})&=&(1+\hat{P}_{12})\left\{
{1\over2}(q_1-q_5)_\sigma-(1+\hat{P}_{23}){q_{1\sigma}\over
D_{\pi^0135}}\left[(q_2,q_3-2q_5)+\right.\right.\nonumber\\
&&\left.\left. a(q_2,q_3-q_5)\left({m^2_\rho\over
D_{\rho35}}-1\right)\right]+{q_{5\sigma}\over
D_{\pi^+123}}\left[-2(q_1,q_2)+a(q_1,q_3-q_2)\times\right.\right.\nonumber\\
&&\left.\left.\left({m^2_\rho\over
D_{\rho23}}-1\right)\right]+(1-\hat{P}_{35})\left[2(q_1-q_5)_\sigma(q_1+q_5,q_2-q_3)-
\right.\right.\nonumber\\&&\left.\left.(q_1+q_5)_\sigma(q_1-q_5,q_2-q_3)\right]
{m^2_\rho\over2D_{\rho15}D_{\rho23}}+{2\over D_{\omega135}}
\left({n_cg^2c_3\over8\pi^2}\right)^2\times\right.
\nonumber\\&&\left.\left[q_{1\sigma}(1-\hat{P}_{35})(k,q_3)(q_2,q_5)+q_{3\sigma}
(1-\hat{P}_{15})(k,q_5)(q_1,q_2)+\right.\right.\nonumber\\
&&\left.\left.q_{5\sigma}(1-\hat{P}_{13})(k,q_1)(q_2,q_3)\right]\left({1\over
D_{\rho13}}+{1\over D_{\rho15}}+{1\over
D_{\rho33}}+\right.\right.\nonumber\\
&&\left.\left.{c_1-c_2-c_3\over2c_3m^2_\rho}\right)\right\},
\label{jplc}\end{eqnarray} where $k=q_1+q_2+q_3+q_5$. The
expression for
$J_\sigma(\rho^-\to\pi^+_{q_1}\pi^-_{q_3}\pi^-_{q_4}\pi^0_{q_5})$
is obtained from Eq.~(\ref{jplc}) by the replacements
$q_1\leftrightarrow q_3$, $q_2\to q_4$ and by inverting an overall
sign. The expression for the contribution of the diagrams
Fig.~\ref{fig7} is
\begin{eqnarray}
M^{(c)}_2&=&{g_{\omega\rho\pi}g_{\rho\pi\pi}\over
f^2_\pi}(1+\hat{P}_{12})(1+\hat{P}_{34})\varepsilon_{\mu\nu\lambda\sigma}q_\mu\epsilon_\nu
\left\{(1+\hat{P}_{24}){q_{1\lambda}q_{3\sigma}\over
D_{\rho13}D_{\pi^0245}}\times\right.\nonumber\\
&&\left.\left[(q_2,q_4-2q_5)+a(q_2,q_4-q_5)\left({m^2_\rho\over
D_{\rho45}}-1\right)\right]-(1-\hat{P}_{13}\hat{P}_{24})\times
\right.\nonumber\\ &&\left.{q_{1\lambda}q_{5\sigma}\over
D_{\rho15}D_{\pi^+234}}\left[-2(q_3,q_4)+a(q_3,q_2-q_4)\left({m^2_\rho\over
D_{\rho24}}-1\right)\right]\right\}.\label{m2c}\end{eqnarray} The
expression for the contribution of the diagrams Fig.~\ref{fig8}
looks as
\begin{eqnarray}
M^{(c)}_3&=&{n_cg\over32\pi^2f^5_\pi}(1+\hat{P}_{12})(1+\hat{P}_{34})
\varepsilon_{\mu\nu\lambda\sigma}q_\mu\epsilon_\nu
\left\{{4c_1-5(c_2+c_3)\over3}\cdot
q_{1\lambda}q_{3\sigma}+\right.\nonumber\\
&&\left.3(c_1-c_2-c_3)\left((1-\hat{P}_{14}\hat{P}_{23}){q_{1\lambda}q_{5\sigma}\over
D_{\pi^+234}}\left[-2(q_3,q_4)+a(q_3,q_2-q_4)\times\right.\right.\right.\nonumber\\
&&\left.\left.\left.\left({m^2_\rho\over
D_{\rho24}}-1\right)\right]-{q_{1\lambda}q_{3\sigma}\over
D_{\pi^0245}}(1+\hat{P}_{24})\left[(q_2,q_4-2q_5)+a(q_2,q_4-q_5)\times\right.\right.\right.
\nonumber\\&&\left.\left.\left.\left({m^2_\rho\over
D_{\rho45}}-1\right)\right] \right) \right\}.
\label{m3c}\end{eqnarray} Again, the relation Eq.~(\ref{relcomp})
is necessary in verifying the Adler condition below. The
expression for the contribution of the diagram Fig.~\ref{fig9} is
\begin{eqnarray}
M^{(c)}_4&=&{g_{\omega\rho\pi}g_{\rho\pi\pi}\over
f^2_\pi}(1+\hat{P}_{12})(1+\hat{P}_{34})\varepsilon_{\mu\nu\lambda\sigma}
\epsilon_\nu\left\{{1\over2} q_\mu\left[{(q_1-q_3)_\lambda
q_{5\sigma}\over
D_{\rho13}}+(1-\hat{P}_{13}\hat{P}_{24})\times\right.\right.\nonumber\\
&&\left.\left.{q_{1\lambda}q_{5\sigma}+{1\over2}(q_1-q_5)_\lambda
q_{2\sigma}\over D_{\rho15}}\right]-{c_1+c_2-c_3\over
4c_3}\left[{q_{5\mu}(q_1-q_3)_\lambda(q_2+q_4)_\sigma\over D_{\rho
13}}+\right.\right.\nonumber\\
&&\left.\left.(1-\hat{P}_{13}\hat{P}_{24}){q_{1\mu}(q_3-q_5)_\lambda
q_{4\sigma}\over D_{\rho35}}\right]\right\}.
\label{m4c}\end{eqnarray} Finally, the amplitude resulting from
the diagram Fig.~\ref{fig10} is
\begin{eqnarray}
M^{(c)}_5&=&-{g_{\omega\rho\pi}g_{\rho\pi\pi}m^2_\rho\over
f^2_\pi}\cdot\left({c_1+c_2-c_3\over
4c_3}\right)(1+\hat{P}_{12})(1+\hat{P}_{34})(1+\hat{P}_{24})\times\nonumber\\
&&\varepsilon_{\mu\nu\lambda\sigma}\epsilon_\nu{(q_1-q_3)_\mu
q_{4\lambda}(q_2-q_5)_\sigma\over D_{\rho13}D_{\rho25}}.
\label{m5c}\end{eqnarray} Notice that the product of the operators
$(1+\hat{P}_{12})(1+\hat{P}_{34})$ makes evident the Bose symmetry
of the full $\omega\to2\pi^+2\pi^-\pi^0$ decay amplitude,
\begin{equation}M(\omega_q\to\pi^+_{q_1}\pi^+_{q_2}\pi^-_{q_3}\pi^-_{q_4}\pi^0_{q_5})=
M^{(c)}_1+M^{(c)}_2+M^{(c)}_3+M^{(c)}_4+M^{(c)}_5.
\label{mfullc}\end{equation}

\subsubsection{Verifying the Adler condition for
the $\omega\to2\pi^+2\pi^-\pi^0$ decay amplitude} \label{subsubc}

Let us write down the Adler limits of all the above contributions
to the $\omega\to2\pi^+2\pi^-\pi^0$ decay amplitudes. As an
example, the case  $q_1=0$ is considered in detail. Representing
the total amplitude Eq.~(\ref{mfullc}) in this limit as
$$M^{(c)}(q_1=0)={g_{\omega\rho\pi}g_{\rho\pi\pi}\over
f^2_\pi}(1+\hat{P}_{34})\varepsilon_{\mu\nu\lambda\sigma}\epsilon_\nu
T_{\mu\lambda\sigma}, $$ one  has the following expressions for
the  diagrams Fig.~\ref{fig6}$-$\ref{fig10}, respectively,
\begin{eqnarray}
T^{(6)}_{\mu\lambda\sigma}(q_1=0)&=&{1\over2}
(1-\hat{P}_{35})q_\mu\left[{q_{4\lambda}(q_2-q_5)_\sigma\over
D_{\rho25}}+{q_{2\lambda}q_{3\sigma}\over
D_{\rho35}}\right],\nonumber\\
T^{(7)}_{\mu\lambda\sigma}(q_1=0)&=& (1-\hat{P}_{35}){q_\mu
q_{2\lambda}q_{3\sigma}\over D_{\rho23}},\nonumber\\
T^{(8)}_{\mu\lambda\sigma}(q_1=0)&=&-\left(2+{c_1+c_2-c_3\over
c_3}\right){q_\mu q_{2\lambda}q_{5\sigma}\over
4m^2_\rho},\nonumber\\
T^{(9)}_{\mu\lambda\sigma}(q_1=0)&=&{1\over6}q_\mu\left[
{-2q_{2\lambda}q_{3\sigma}+(q_2-q_3)_\lambda(2q_5-q_4)_\sigma\over
D_{\rho35}}-\right.\nonumber\\&&\left.{4q_{3\lambda}q_{5\sigma}+
(q_3-q_5)_\lambda(2q_4-q_2)_\sigma\over
D_{\rho23}}+q_{2\lambda}q_{5\sigma}\left({3\over
D_{\rho25}}+{3\over m^2_\rho}\right)\right]+\nonumber\\&&{c_1+c_2-c_3\over4c_3}
\left[{q_\mu q_{2\lambda}q_{5\sigma}\over
m^2_\rho}+(1-\hat{P}_{24}\hat{P}_{35}){q_{3\mu}(q_4-q_5)_\lambda
q_{2\sigma}\over
D_{\rho45}}\right],\nonumber\\T^{(10)}_{\mu\lambda\sigma}(q_1=0)&=&
-{c_1+c_2-c_3\over4c_3}(1-\hat{P}_{24}\hat{P}_{35}){q_{3\mu}(q_4-q_5)_\lambda
q_{2\sigma}\over D_{\rho45}}. \label{adlerc1}\end{eqnarray} In the above formulas, the
upper index points to the label of the corresponding figure. Once again,
when obtaining $T^{(8)}_{\mu\lambda\sigma}$, the relation
(\ref{relcomp}) is essential. The close inspection of
Eq.~(\ref{adlerc1}) shows that the sum of above tensors vanishes.
Indeed, the cancellation of  the $\rho$ pole terms
proportional to $c_1+c_2-c_3$ and of all the non-$\rho$-pole terms  is evident.
Let us check the cancellation of the $\rho$ pole terms taking as
an example the terms $\propto q_\mu/6D_{\rho25}$. One has for the
sum of such terms the expression
$3[q_{4\lambda}(q_2-q_5)_\sigma-q_{2\lambda}q_{5\sigma}]$.
Applying the operator $1+\hat{P}_{34}$, taking into
account the four-momentum conservation and the presence of the tensor
$\epsilon_{\mu\nu\lambda\sigma}$ one finds that the above
combination vanishes. The cancellation of the remaining $\rho$
pole terms is checked in the same manner.
The cases of $q_{2,3,4}=0$ are obtained from the
present case by Bose symmetry and by the evident replacements of the pion
momenta. In the case $q_5=0$, the contributions of the diagrams
Fig.~\ref{fig8}, \ref{fig9}, and \ref{fig10} vanish  separately, while the contributions of the diagrams
Fig.~\ref{fig6} and \ref{fig7} are equal in magnitude but opposite
in sign, hence they are cancelled. The conditions of the vanishing
of the amplitude in the Adler limit obtained in subsection
\ref{subsubn} and this subsection turn out to be of great
importance in obtaining the $\phi\to5\pi$ decay amplitudes.

\section{The $\omega\to\pi^+\pi^-3\pi^0$ and
$\omega\to2\pi^+2\pi^-\pi^0$ branching ratios
revisited.}\label{omegaresults}

\begin{figure}
\includegraphics{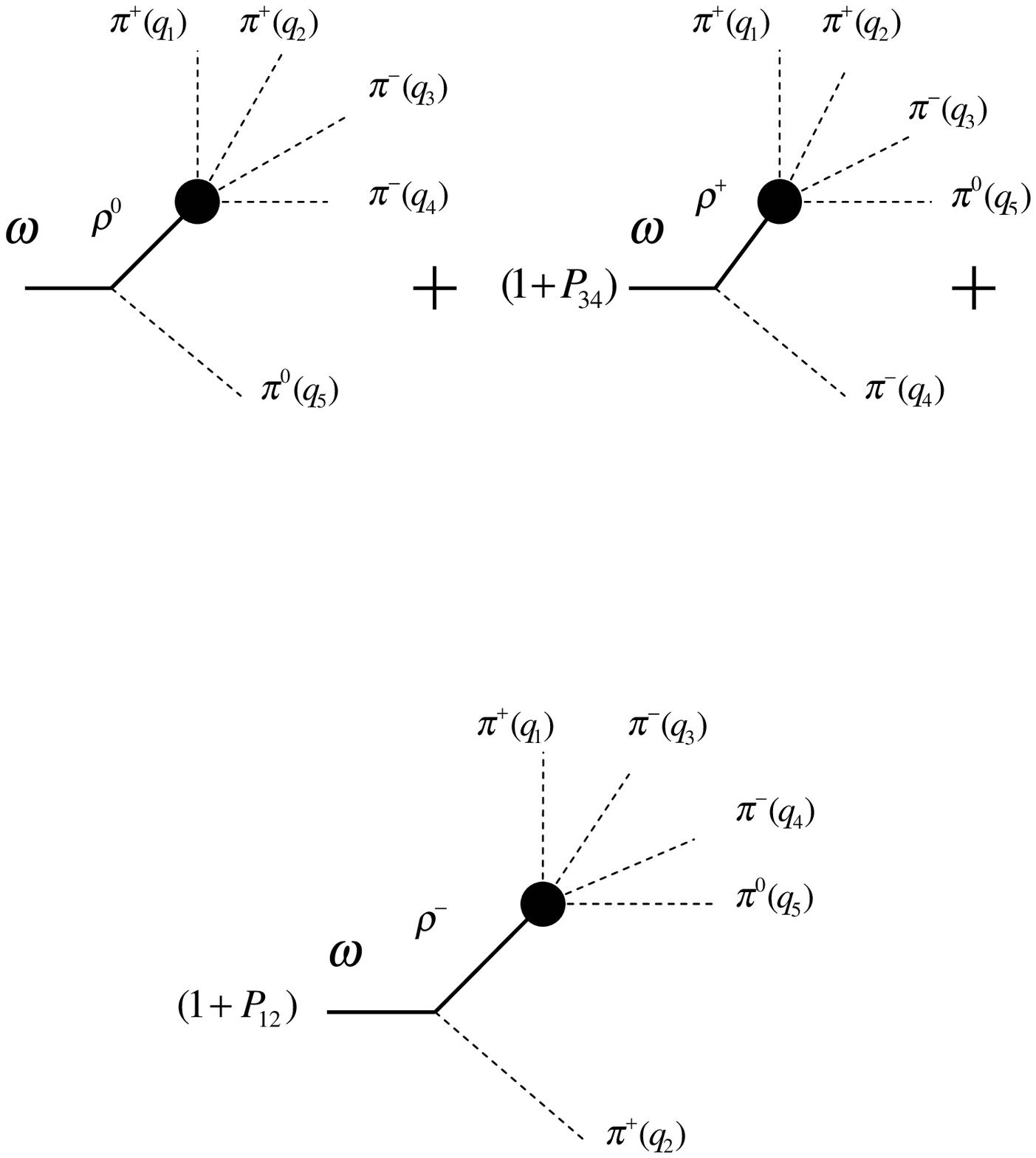}
\caption{\label{fig6}The diagrams describing the amplitudes of the decay
$\omega\to2\pi^+2\pi^-\pi^0$ through the $\rho$ intermediate state
followed by the decay $\rho\to4\pi$. The shaded circles refer to
the whole $\rho\to4\pi$ amplitudes.}
\end{figure}

\begin{figure}
\includegraphics{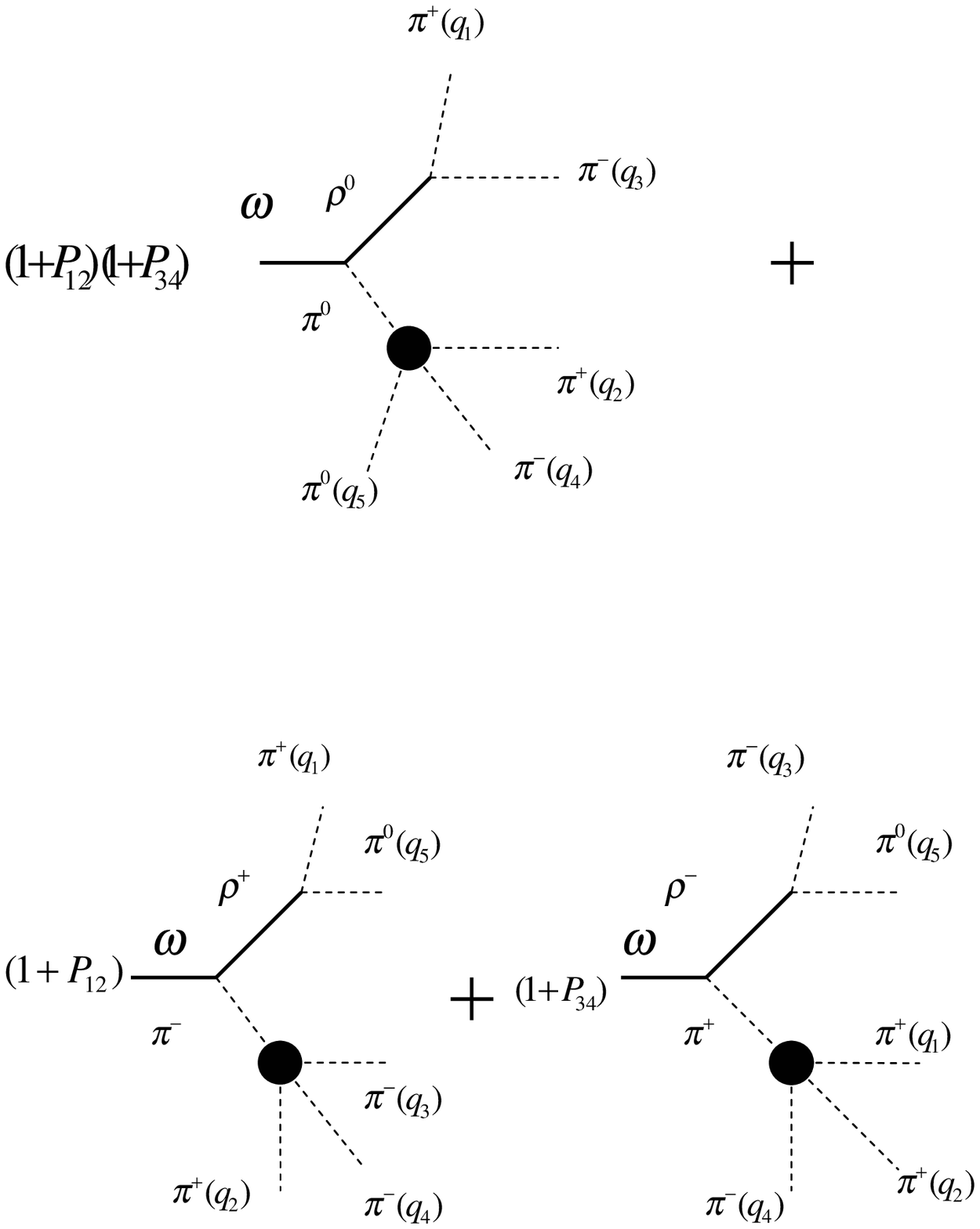}
\caption{\label{fig7}The diagrams describing the amplitudes of the decay
$\omega\to2\pi^+2\pi^-\pi^0$ through the $\rho\pi$ intermediate
state followed by the transitions $\rho\to2\pi$ and $\pi\to3\pi$.
The shaded circles refer to the effective $\pi\to3\pi$ vertices
given by Eq.~(\protect\ref{4pi}). Note that non-$\pi$-pole term is
included to the first pair of diagrams in Fig.~\protect\ref{fig9}
below.}
\end{figure}

\begin{figure}
\includegraphics{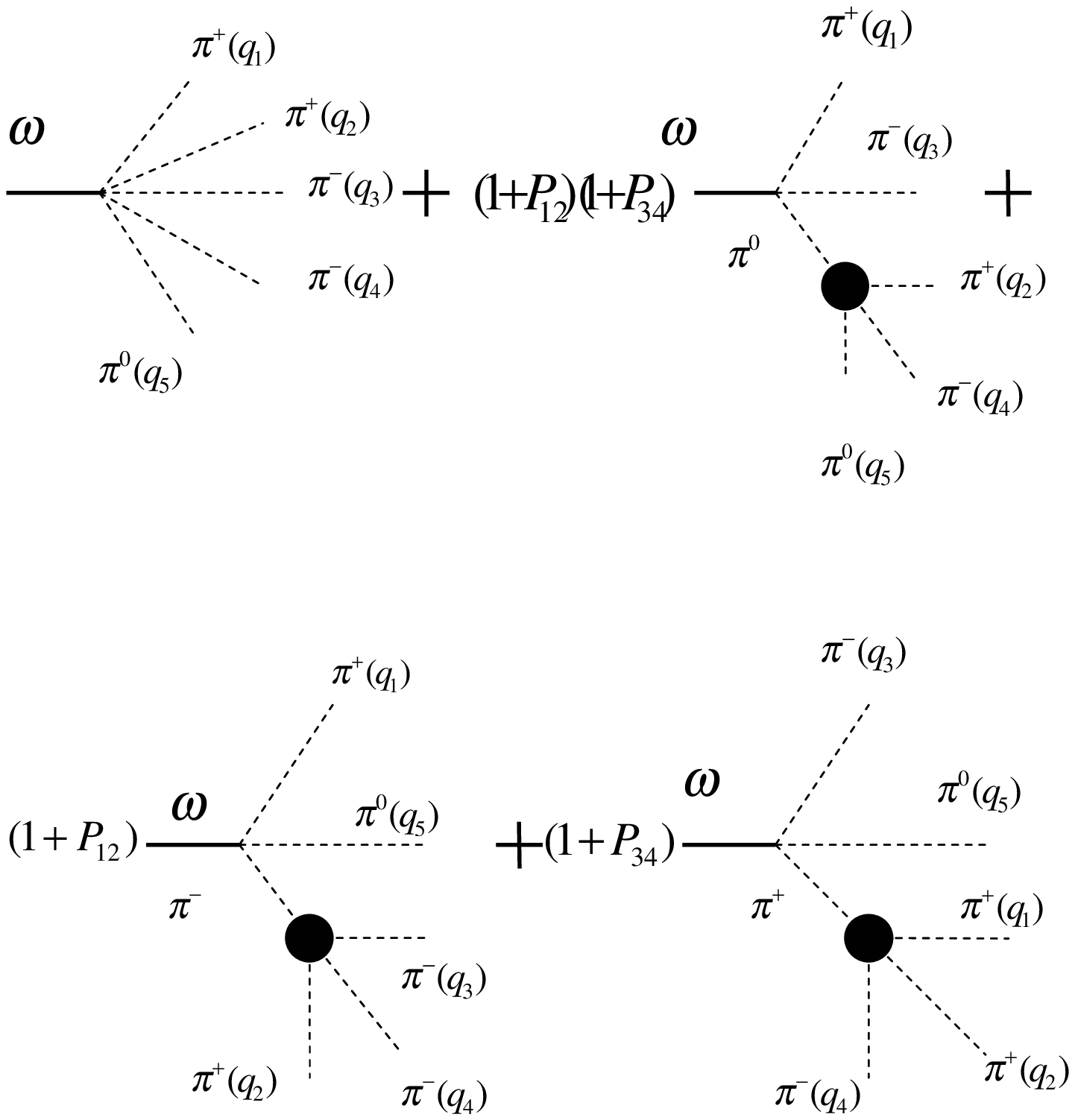}
\caption{\label{fig8}The diagrams describing the contributions to the
$\omega\to2\pi^+2\pi^-\pi^0$ decay amplitude via point like
vertices. The shaded circles refer to the effective $\pi\to3\pi$
vertices given by Eq.~(\protect\ref{4pi}).}
\end{figure}

\begin{figure}
\includegraphics{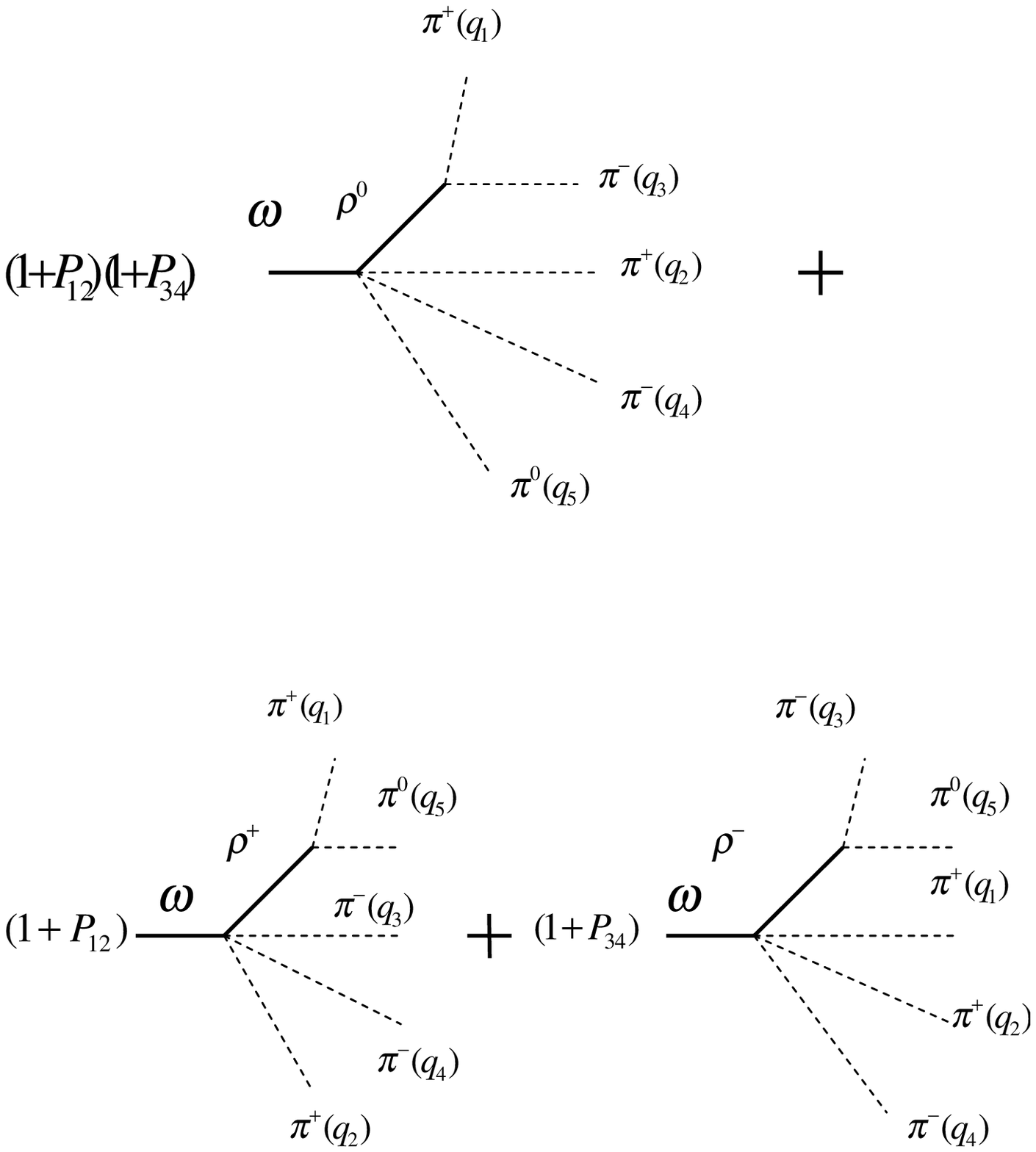}
\caption{\label{fig9}The contributions to the $\omega\to2\pi^+2\pi^-\pi^0$ decay amplitude
arising due to the chiral $\omega\to\rho3\pi$ vertex.}
\end{figure}

\begin{figure}
\includegraphics{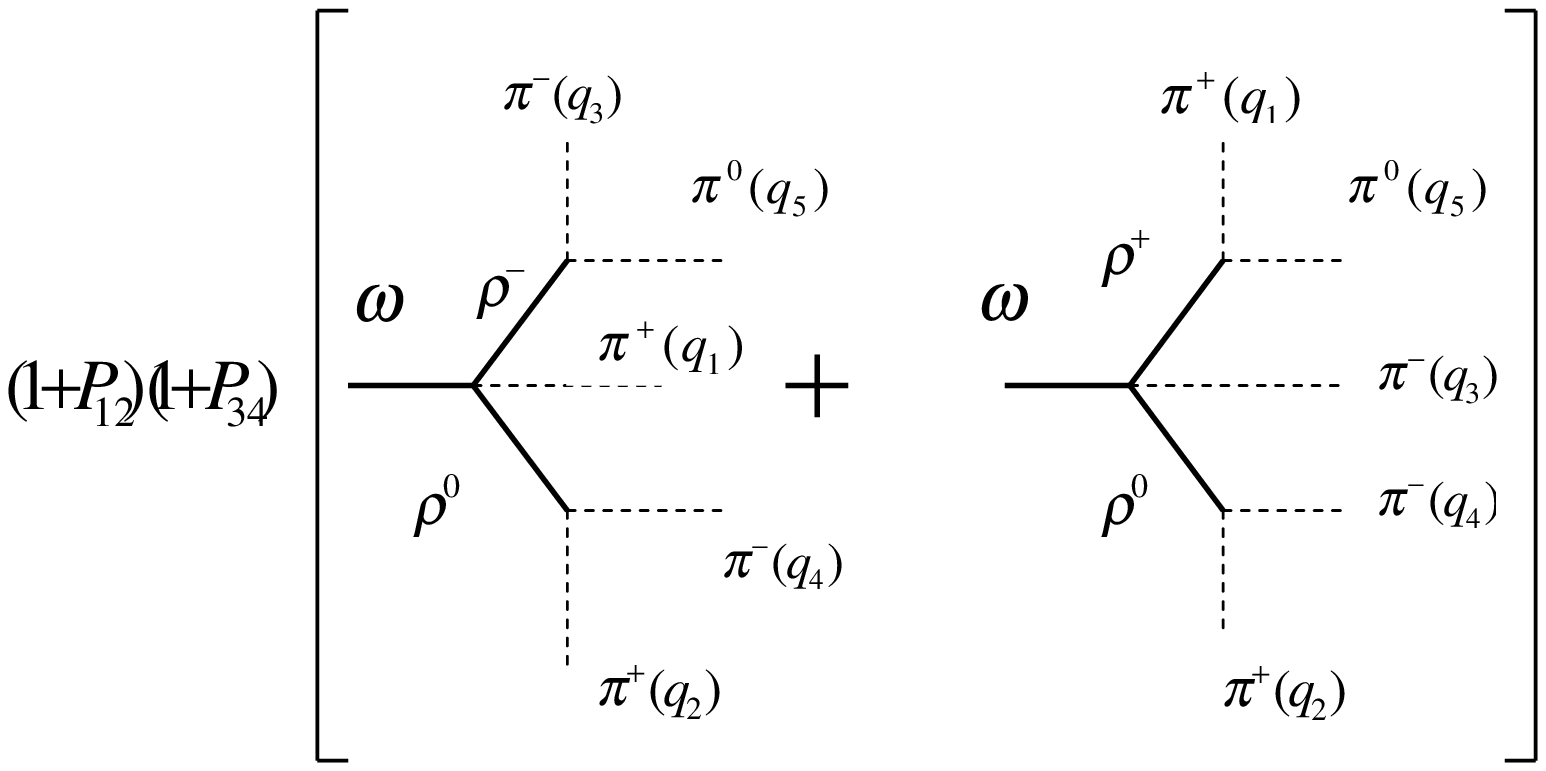}
\caption{\label{fig10}The contributions to the $\omega\to2\pi^+2\pi^-\pi^0$ decay amplitude
via intermediate state with two $\rho$ mesons.}
\end{figure}

In our previous work Ref.~\cite{ach00}, the branching ratios of
the $\omega\to\pi^+\pi^-3\pi^0$ and $\omega\to2\pi^+\pi^-\pi^0$
decays were estimated. Essential for that evaluation were the
expressions for the contributions of the diagrams shown in
Fig.~\ref{fig1} and  \ref{fig6}, added with the specific
correction factor stemming from the diagrams shown in
Fig.~\ref{fig2} and \ref{fig7} of the present paper, respectively.
This seemed to be justifiable because of the presence of the
$\rho$ pole. As it will become clear later on, the non-$\rho$-pole
terms are essential.

Strictly speaking, the HLS approach does not give the {\it
predictions} even for the $\omega\to\pi^+\pi^-\pi^0$ decay rate,
because arbitrary constants $c_{1,2,3}$ enter the expression for
Lagrangian Eq.~(\ref{lan}). As was pointed out in
Ref.~\cite{hls1,hls2}, these constants should be determined from
experiment. Nevertheless, HLS relates the contributions to the
amplitudes, compare Fig.~\ref{fig1}, \ref{fig2} to
Fig.~\ref{fig3}, \ref{fig4}, and \ref{fig5} (respectively,
Fig.~\ref{fig6}, \ref{fig7} to Fig.~\ref{fig8}, \ref{fig9}, and
\ref{fig10}), which otherwise appear unrelated. One can obtain
reasonable predictions for the $\omega\to5\pi$ decay rates upon
assuming particular relations among $c_{1,2,3}$. First, there are
no experimental indications on the point like
$\omega\to\pi^+\pi^-\pi^0$ vertex, hence one can take the relation
Eq.~(\ref{rel1}) for granted.  Second, the constant $c_3$, see
Eq.~(\ref{c3}), extracted from the $\omega\to3\pi$ branching
ratio, is remarkably close to unity. Note that older chiral models
\cite{schechter} for the vector mesons interactions, added with
the terms arising from the gauging the anomalous Wess-Zumino
action \cite{wza}, predicted $c_3=1$. We fix $c_3$ from the
$\omega\to3\pi$ partial width, see Eq.~(\ref{gomrp}) and
(\ref{c3}). After taking into account Eq.~(\ref{rel1}), the ratio
$c_1/c_3$ remains arbitrary, and the magnitude of the
$\omega\to5\pi$ decay width depends on this parameter. We choose
its value guided by the following considerations. The inspection
of the expressions for the $\omega\to5\pi$ decay amplitudes
obtained in Sec.~\ref{omegaamp} shows that almost all the terms
except those proportional to $c_1+c_2-c_3$, has the tensor
structure
\begin{equation}
M={g_{\omega\rho\pi}g_{\rho\pi\pi}\over f^2_\pi}
\varepsilon_{\mu\nu\lambda\sigma}q_\mu\epsilon_\nu
T_{\lambda\sigma},\label{ampstruct}\end{equation} where
\begin{equation}
T_{\lambda\sigma}=\sum_{a<b}G_{ab}q_{(a)\lambda}q_{(b)\sigma}\label{T}\end{equation}
is the tensor composed of pion four-momenta $q_{(a)}$, $a=1,...5$,
and $G_{ab}$ are invariant amplitudes, whose explicit form can be
read off the expressions for the amplitudes obtained in
Sec.~\ref{omegaamp} by gathering the coefficients in front of
$q_{(a)\lambda}q_{(b)\sigma}$. They are very lengthy, so we do not
give them here. In the rest frame system of the decaying $\omega$,
the Lorentz structure of Eq.~(\ref{ampstruct}) is reduced to the
three dimensional form $e_{ijk}\xi_iT_{jk}$, where $\bm{\xi}$ is
the polarization vector of the $\omega$ in this frame, $e_{ijk}$
is totally antisymmetric in $i,j,k=1,2,3$. This enormously
simplify the calculation of the modulus squared of the amplitude.
In the meantime, the terms proportional to $c_1+c_2-c_3$ has
entirely the  four dimensional tensor structure
$\varepsilon_{\mu\nu\lambda\sigma}\epsilon_\mu
q_{(a)\nu}q_{(b)\lambda}q_{(c)\sigma}$. The resulting expression
for the modulus squared of the full amplitude turns out to be
extremely lengthy. For the sake of simplicity, we set
\begin{equation}
c_1+c_2-c_3=0\label{rel2}\end{equation} in what follows.
Note that this means that the contributions of the diagrams Fig.~\ref{fig5},
\ref{fig10} together with the part of the contributions from the
diagrams Fig.~\ref{fig4} and \ref{fig9} are dropped.
The results of relaxing the condition Eq.~(\ref{rel2}) are discussed at the end
of the present Section. Finally, our
assumptions about HLS arbitrary constants $c_{1,2,3}$ and $a$ are
\begin{equation}
c_1=c_3\mbox{, }c_2=0\mbox{,
}a=2.\label{sumrel}\end{equation}Notice that the above relations
among $c_{1,2,3}$ are the solutions of Eq.~(\ref{rel1}) and
(\ref{rel2}).

The expression for the partial width of the decays (\ref{reacn})
and (\ref{reacc})  looks as
\begin{equation}
\Gamma_{\omega\to5\pi}(s)={1\over2\sqrt{s}(2\pi)^{11}N_{\rm sym}
}\int|M|^2d{\cal D}_5,\label{gamma}\end{equation} where
$s=\left(\sum_{a=1}^5q_a\right)^2$ is the total energy squared in
the rest frame system of the decaying particle, the Bose symmetry
factor $N_{\rm sym}=6\mbox{, }4$ for the reaction (\ref{reacn})
and (\ref{reacc}), respectively, and $d{\cal D}_5$ given in
Ref.~\cite{kumar} is the differential element of the phase space
volume of the five pion final state. Note that we take into
account  the mass difference of the charged and neutral pions both
in amplitude and in the phase space volume. In the above formula,
\begin{equation}
|M|^2={1\over3}\left({g_{\omega\rho\pi}g_{\rho\pi\pi}\over
f^2_\pi}\right)^2{s\over2}\sum_{i,j=1}^3|T_{ij}-T_{ji}|^2\label{msquar}\end{equation}
is the modulus squared of the amplitude Eq.~(\ref{ampstruct})
averaged over three independent polarizations of the $\omega$.
When evaluating Eq.~(\ref{gamma}), eight Mandelstam-like invariant
variables $s_i$, $u_i$, $i=1,2,3$, and $t_1$, $t_2$ proposed by
Kumar in Ref.~\cite{kumar} are suitable. They are given in
Appendix. All the scalar products of the pair of pion four-momenta
are expressed via the Kumar variables by the expressions given in
Appendix. For the numerical evaluation of the eight dimensional
integral over Kumar variables we use the method suggested in
Ref.~\cite{sag}.

We evaluate both the branching ratios for the two mentioned
isotopic modes at the resonance mass,
\begin{equation}
B_{\omega\to5\pi}(m^2_\omega)={\Gamma_{\omega\to5\pi}(m^2_\omega)\over\Gamma_\omega},
\label{brpoint}\end{equation} and the branching ratios averaged
over resonance peak,
\begin{equation}
B^{\rm
aver}_{\omega\to5\pi}={2\over\pi}\int_{\sqrt{s}=m_\omega-\Gamma_\omega}
^{\sqrt{s}=m_\omega+\Gamma_\omega}d\sqrt{s}{s\Gamma_{\omega\to5\pi}(s)\over(s-m^2_\omega)^2
+m^2_\omega\Gamma^2_\omega}.\label{braver}\end{equation}The
quantity $B^{\rm aver}_{\omega\to5\pi}$ is useful in situations
where the total energy of the five pion state is not directly
measured, as is the case in, e.g., photoproduction or peripheral
production in $\pi N$ collisions. The results of the evaluation
are the following:
\begin{eqnarray}
B_{\omega\to\pi^+\pi^-3\pi^0}(m^2_\omega)&=&3.6\times10^{-9}\mbox{,
}B^{\rm
aver}_{\omega\to\pi^+\pi^-3\pi^0}=2.8\times10^{-9},\nonumber\\
B_{\omega\to2\pi^+2\pi^-\pi^0}(m^2_\omega)&=&3.5\times10^{-9}\mbox{,
}B^{\rm aver}_{\omega\to2\pi^+2\pi^-\pi^0}=2.7\times10^{-9}.
\label{bromega}\end{eqnarray} These branching ratios for the
$\omega\to5\pi$ decay   by the factor of more than three exceed
those obtained in our previous paper Ref.~\cite{ach00}. The reason
of the disagreement is the following. As is mentioned in the
beginning of the present Section, the diagrams Fig.~\ref{fig1} and
\ref{fig6} corrected with those of Fig.~\ref{fig2} and \ref{fig7}
were considered to be dominant in Ref.~\cite{ach00}. Let us
evaluate the contributions of the diagrams Fig.~\ref{fig1} and
\ref{fig6} to the branching ratios of the decays
$\omega\to\pi^+\pi^-3\pi^0$ and $\omega\to2\pi^+2\pi^-\pi^0$,
respectively. By the reason soon to become clear in the case of
the $\phi\to5\pi$ decay, we call these contributions resonant. One
obtains  $B^{\rm resonant}
_{\omega\to\pi^+\pi^-3\pi^0}=1.54\times10^{-9}$,  and $B^{\rm
resonant} _{\omega\to2\pi^+2\pi^-\pi^0}=1.4\times10^{-9}$. These
figures are close to $B_{\omega\to\pi^+\pi^-3\pi^0}\simeq
B_{\omega\to2\pi^+\pi^-\pi^0} \simeq1\times10^{-9}$ obtained in
Ref.~\cite{ach00}. The  evaluation of the net contribution of all
the remaining diagrams called non-resonant gives $B^{\rm
non-resonant}_{\omega\to\pi^+\pi^-3\pi^0}=0.47\times10^{-9}$ and
$B^{\rm
non-resonant}_{\omega\to\pi^+\pi^-3\pi^0}=0.50\times10^{-9}$. The
non-resonant contributions  amount to 13-14 \% of the total ones
Eq.~(\ref{bromega}). However, the phase space averaged relative
phase  difference between the resonant and non-resonant
contributions evaluated with the  above numbers is
$\delta=21^\circ$, and $\delta=17^\circ$, respectively, for the
reaction (\ref{reacn}) and (\ref{reacc}). These phase differences
and the comparison with the total branching ratios
Eq.~(\ref{bromega}) show that the mentioned contributions to the
decay amplitude are almost in phase. The neglect of seemingly
small non-resonant contributions resulted in the underestimated
magnitude of branching ratios in Ref.~\cite{ach00}.

\begin{figure}
\includegraphics{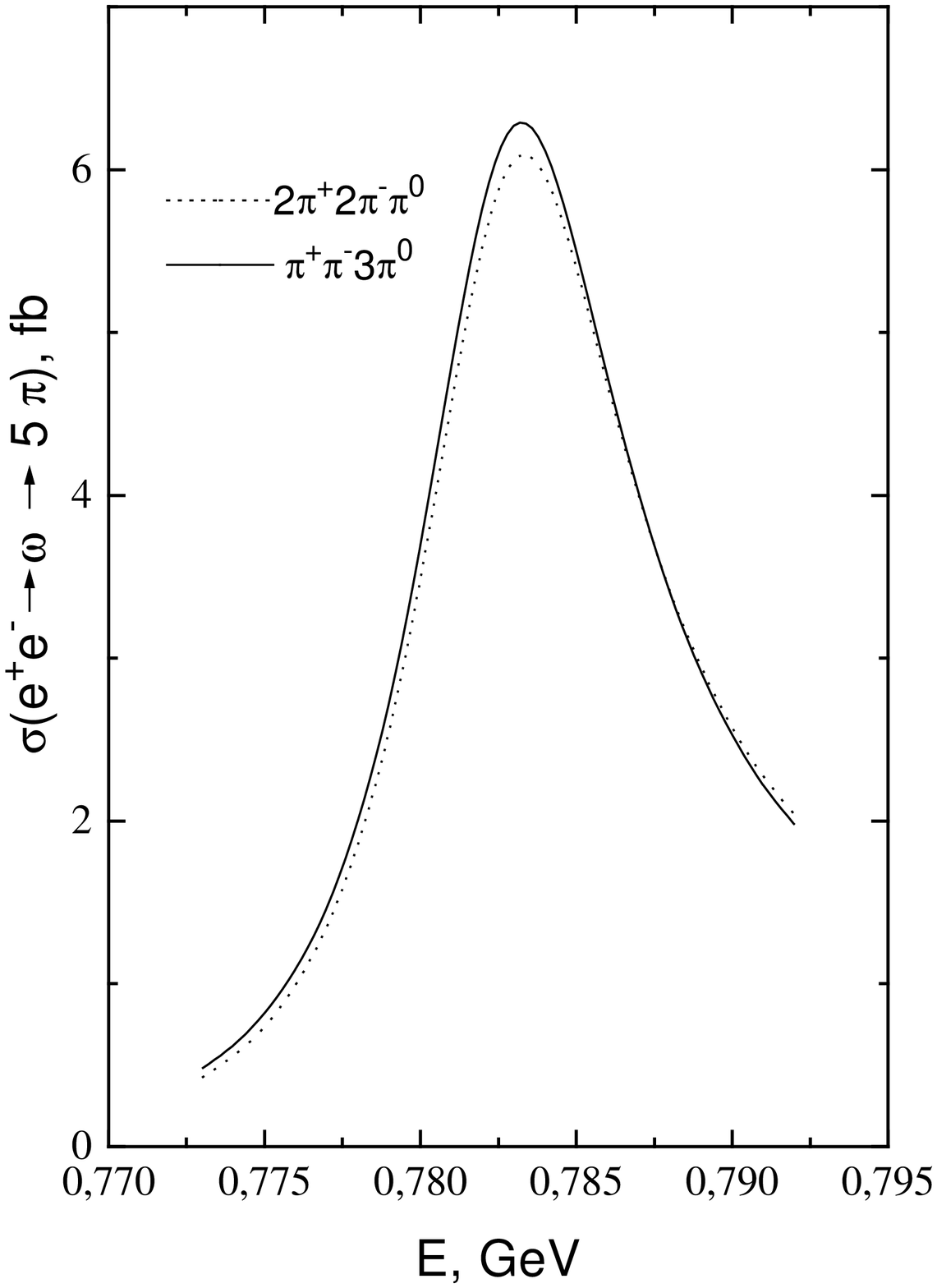}
\caption{\label{fig11}The excitation curve of the decays  $\omega\to5\pi$ in $e^+e^-$
annihilation.}
\end{figure}

The excitation curves for the $\omega\to5\pi$ decays in $e^+e^-$
annihilation,
\begin{equation}
\sigma_{\omega\to5\pi}(s)=12\pi\left({m_\omega\over\sqrt{s}}\right)^3\Gamma_{\omega\to
e^+e^-}(m^2_\omega){s\Gamma_{\omega\to5\pi}(s)\over(s-m^2_\omega)^2
+m^2_\omega\Gamma^2_\omega}\label{sigomega}\end{equation} are
plotted in Fig.~\ref{fig11}. The curves are asymmetric and shifted
by 0.7 MeV towards the  higher values from the $\omega$ mass because of strong energy
dependence of $\Gamma_{\omega\to5\pi}(s)$, see Fig.~\ref{fig12}
below. As is seen, both isotopic channels have approximately equal
branching ratios and almost coincident excitation curves in the
$\omega$ resonance region. This can be understood as follows. The
matrix elements squared numerically are approximately the same in
the near-to-threshold region, since the pion mass difference is
smeared  in the sum of various contributions. Hence, they are
cancelled in the ratio of two partial widths, leaving the ratio of
the phase space volumes. Using the nonrelativistic expression for
the phase space volume of the five pion final state from
Ref.~\cite{byck}, one obtains\begin{equation}
{B(\omega\to2\pi^+2\pi^-\pi^0,m^2_\omega)\over
B(\omega\to\pi^+\pi^-3\pi^0,m^2_\omega)}={3m_{\pi^+}\over2m_{\pi^0}}
\left({2m_{\pi^+}+3m_{\pi^0}\over4m_{\pi^+}+m_{\pi^0}}\right)^{3/2}
\left({m_\omega-4m_{\pi^+}-m_{\pi^0}\over
m_\omega-2m_{\pi^+}-3m_{\pi^0}}\right)^5=0.93\label{omegarat}\end{equation}
to be compared to 0.96 calculated from Eq.~(\ref{bromega}). The
ratio of the Bose symmetry factors 3/2 compensates the smaller
phase space volume of the final state $2\pi^+2\pi^-\pi^0$ as
compared to $\pi^+\pi^-3\pi^0$ one. In the meantime, the energy
dependence of the $\omega\to5\pi$ partial width in the dynamical
model is drastically different from that in the model of the
Lorentz-invariant phase space (lips). In the latter, one has the
following expression for the $\omega\to5\pi$ partial width:
\begin{equation}
\Gamma_{\omega\to5\pi}^{(\rm
lips)}(s)=\Gamma_{\omega\to5\pi}(m^2_\omega){W_{5\pi}(s)\over
W_{5\pi}(m^2_\omega)},\label{widps}\end{equation} where
$\Gamma_{\omega\to5\pi}(m^2_\omega)$ is the partial width
evaluated with the dynamical amplitudes given in
Sec.~\ref{omegaamp}, and the expression for the Lorentz invariant
phase space volume  is
\begin{eqnarray}
W_{5\pi}(s)&=&{\pi^4\over(2\pi)^{11}32s^{3/2}N_{\rm sym}}
\int_{(m_1+m_2+m_3+m_4)^2}^{(\sqrt{s}-m_5)^2 } {ds_1\over
s_1}\lambda^{1/2}(s,s_1,m^2_5)\int_{(m_1+m_2+m_3)^2}^{(\sqrt{s_1}-m_4)^2}
{ds_2\over s_2}\times\nonumber\\
&&\lambda^{1/2}(s_1,s_2,m^2_4)\int_{(m_1+m_2)^2}^{(\sqrt{s_2}-m_3)^2}
{ds_3\over
s_3}\lambda^{1/2}(s_2,s_3,m^2_3)\lambda^{1/2}(s_3,m^2_1,m^2_2),\label{w5pi}\end{eqnarray}
with $m_i$, $i=1,...5$ being the mass of the meson $\pi_i$, and
\begin{equation}
\lambda(x,y,z)=x^2+y^2+z^2-2xy-2xz-2yz.\label{lambda}\end{equation}
The predictions of both models for the energy dependence of
$\Gamma_{\omega\to2\pi^+2\pi^-\pi^0}(s)$ are plotted in
Fig.~\ref{fig12}. The plot for the $\pi^+\pi^-3\pi^0$ final state
looks similar and is not given here. The faster growth of the
partial width in the dynamical model as compared to the phase
space one is due to the resonance enhancement arising from opening
of the $\rho$ production in the intermediate state.

Let us relax the constraint Eq.~(\ref{rel2}) on the parameters
$c_{1,2,3}$. To be specific, we choose $-2\leq c_2\leq2$ instead of
$c_2=0$ assumed earlier. The
corresponding  ratio $\gamma=(c_1+c_2-c_3)/4c_3$ parametrizing
the strength of the neglected terms then falls into the interval
$-1\leq\gamma\leq1$, or $-1\leq c_1/c_3\leq3$. The  branching ratios
$B_{\omega\to5\pi}$ evaluated with the new parameters deviate  by less
than one percent from those evaluated at $\gamma=0$.

\section{The evaluation of the  $\phi\to\pi^+\pi^-3\pi^0$ and
$\phi\to2\pi^+2\pi^-\pi^0$ branching ratios.}\label{phiresults}

As is known, chiral models, including HLS, do not possess the
terms responsible for the decays of $\phi$ meson into final states
containing nonstrange quarks only. However, one can guess the
general form of such terms guided by both the OZI rule violation
in the decay $\phi\to\rho\pi\to\pi^+\pi^-\pi^0$ and by the
Adler condition.

There are two feasible models of the OZI-suppressed
$\phi\to\rho\pi$ decay amplitude. The first one is the
$\phi\omega$ mixing model, where the above decay proceeds due to
the small admixture of nonstrange quarks in the flavor wave
function of $\phi$ meson composed mostly of the pair of strange
quarks. In the second model $\phi$ goes to $\rho\pi$ directly,
see Ref.~\cite{ach95}. Earlier we pointed out
that there are no particular reasons to prefer one model to
another, and possible ways to resolve the issue were pointed out
\cite{ach95,ach93}. Recent SND data \cite{snd03} point to a
sizeable coupling constant of direct $\phi\to\rho\pi$ transition,
assuming the dependence $|\psi(0,m_V)|^2\propto m^2_V$\cite{ach95}
of the wave function of the vector $q\bar q$ bound state at the
origin on the mass $m_V$ of this state. It should be noted that
the assumed dependence agrees remarkably good with the ratios of
the measured leptonic widths of the vector quarkonia $\rho$,
$\omega$, $\phi$, $J/\psi$, and $\Upsilon(1S)$.

The decays $\phi\to5\pi$ are treated slightly differently in the
above models of  OZI rule violation. Let us consider them in due turn.
In the model of $\phi\omega$ mixing $\phi$ goes to the
off-mass-shell $\omega$ which decays as is considered in
Sec.~\ref{omegaamp}. Hence, one can immediately obtain
\begin{equation}
\Gamma_{\phi\to5\pi}(m^2_\phi)=|\varepsilon_{\phi\omega}(m^2_\phi)|^2
\Gamma_{\omega\to5\pi}(m^2_\phi),\label{gamphimix}\end{equation}
where $\varepsilon_{\phi\omega}(m_\phi)$ is the complex parameter
of $\phi\omega$ mixing taken at the $\phi$ mass. It can be
evaluated as
$$|\varepsilon_{\phi\omega}(m^2_\phi)|^2={\Gamma_{\phi\to3\pi}(m^2_\phi)\over
\Gamma_{\omega\to3\pi}(m^2_\omega)}\cdot r=3.04\times10^{-3},$$
where $r=3.5\times10^{-2}$ is the ratio of the three pion phase
space volumes at the $\omega$ and $\phi$ peaks.

If $\phi\omega$ mixing is negligible, one should introduce a
number of new OZI rule violating parameters to quantify the
$\phi\to5\pi$ decay amplitude. Guided by the condition of chiral
symmetry expressed as the demand that the correct decay amplitude
should fulfill the Adler condition, it is reasonable to expect
that the effective Lagrangian describing anomalous OZI suppressed
decays of $\phi$ meson looks similar to the Lagrangian
Eq.~(\ref{lan}),
\begin{eqnarray}
\cal{L}^{\rm
an}_{\phi,\rho,\pi}&=&
{1\over2f^3_\pi}(\beta_1-\beta_2-\beta_3)\varepsilon_{\mu\nu\lambda\sigma}\phi_\mu
\left(\partial_\nu\bm{\pi}\cdot[\partial_\lambda\bm{\pi}\times\partial_\sigma\bm{\pi}]
\right)+        \nonumber\\ &&
{1\over8f^5_\pi}\left[-\beta_1+{5\over3}\left(\beta_2+\beta_3\right)\right]
\varepsilon_{\mu\nu\lambda\sigma}\phi_\mu
\left(\partial_\nu\bm{\pi}\cdot[\partial_\lambda\bm{\pi}\times\partial_\sigma\bm{\pi}]
\right)\bm{\pi}^2-        \nonumber\\ && {2\beta_3g\over
f_\pi}\varepsilon_{\mu\nu\lambda\sigma}\partial_\mu\phi_\nu
\left\{\left(\bm{\rho}_\lambda\cdot\partial_\sigma\bm{\pi}\right)+{1\over6f^2_\pi}
\left[\left(\bm{\rho}_\lambda\cdot\bm{\pi}\right)\left(\bm{\pi}\cdot\partial_\sigma
\bm{\pi}\right)-\bm{\pi}^2\left(\bm{\rho}_\lambda\cdot\partial_\sigma\bm{\pi}\right)
\right]\right\}-     \nonumber\\ && {2g\over
f_\pi}(\beta_1+\beta_2-\beta_3)\varepsilon_{\mu\nu\lambda\sigma}\phi_\mu
\left\{\frac{1}{4f^2_\pi}\left(\partial_\nu\bm{\pi}\cdot\bm{\rho}_\lambda\right)\left(\bm{\pi}\cdot
\partial_\sigma\bm{\pi}\right)-{g\over4}\left(\left[\bm{\rho}_\nu\times
\bm{\rho}_\lambda\right]\cdot\partial_\sigma\bm{\pi}\right)\right\},
\label{lanphi}
\end{eqnarray}
where $\beta_{1,2,3}$ are the above mentioned parameters
responsible for the violation of the OZI rule in the $\phi\to5\pi$
decays of $\phi$ meson. The analysis  similar to that presented
in Sec.~\ref{subsubn} and \ref{subsubc} shows that the
$\phi\to5\pi$ decay amplitudes obtained from the  Lagrangian
(\ref{lanphi}), satisfy the Adler condition. As is evident from
Eq.~(\ref{lanphi}), one should identify the coupling constant of
direct $\phi\to\rho\pi$ transition as
\begin{equation}
g_{\phi\rho\pi}=-{2\beta_3g\over f_\pi}=0.8\mbox{ GeV}^{-1}
,\label{gphirp}\end{equation}where the magnitude of
$g_{\phi\rho\pi}$ is obtained from the $\phi\to3\pi$ partial
widths, while the positive sign (relative to $g_{\omega\rho\pi}$
usually taken to be positive) is fixed by the $\phi\omega$
interference pattern observed in the energy dependence of the
$e^+e^-\to\pi^+\pi^-\pi^0$ reaction cross section \cite{snd01}.
Note that we neglect the unitarity corrections to
$g_{\phi\rho\pi}$ \cite{ach00a},  because they are irrelevant in
the context of the present work. Next, it seems to be  no sizeable
point like $\phi\to\pi^+\pi^-\pi^0$ contribution. Indeed, first,
the existing upper limit to the branching ratio of the
non-$\rho\pi$ intermediate state direct transition
$\phi\to\pi^+\pi^-\pi^0$ obtained by SND group at VEPP-2M, is very
small \cite{snd02},
\begin{equation}B^{\rm
direct}(\phi\to\pi^+\pi^-\pi^0)<6\times10^{-4}(\mbox{90 \%
C.L.}).\label{bound3pi}\end{equation}  Second, the KLOE
Collaboration at DA$\Phi$NE gives the phase space averaged direct
$\phi\to\pi^+\pi^-\pi^0$ contribution at the level of 1\% \cite{kloe}
of the total $\pi^+\pi^-\pi^)$ decay rate. Hence, in a close analogy
with the $\omega$ case, one can set
\begin{equation}\beta_1-\beta_2-\beta_3=0.\label{betcond1}\end{equation}
The results of relaxing this conditions are discussed at the end
of the present Section.
The magnitude  $\beta_3=-0.006$ is fixed according to Eq.~(\ref{gphirp}) by
the $\phi\to3\pi$ partial widths. After all, the ratio
$\beta_1/\beta_3$ remains arbitrary. We set
\begin{equation}
\beta_1+\beta_2-\beta_3=0,\label{betcond2}\end{equation} hence $\beta_1=\beta_3\mbox{,
}\beta_2=0$, so that the $\phi\to5\pi$ decay amplitudes are
determined by the only parameter $\beta_3$ and looks like
Eq.~(\ref{ampstruct}) for the $\omega\to5\pi$ decay, with the
replacement $g_{\omega\rho\pi}\to g_{\phi\rho\pi}$. The tensor
$T_{\lambda\sigma}$ is the same as in the $\omega\to5\pi$ decay
amplitude. Under these assumptions both mentioned models for the
OZI rule violating decay $\phi\to3\pi$ give similar results for
branching ratios of the decays $\phi\to5\pi$. These are the
following:
\begin{eqnarray}
B_{\phi\to\pi^+\pi^-3\pi^0}(m^2_\phi)&=&2.4\times10^{-7}\mbox{,
}B^{\rm
aver}_{\phi\to\pi^+\pi^-3\pi^0}=1.8\times10^{-7},\nonumber\\
B_{\phi\to2\pi^+2\pi^-\pi^0}(m^2_\phi)&=&6.9\times10^{-7}\mbox{,
}B^{\rm aver}_{\phi\to2\pi^+2\pi^-\pi^0}=4.9\times10^{-7},
\label{brphi}\end{eqnarray}where $B^{\rm aver}$, useful for the
reactions of peripheral production, stands for the branching ratio
averaged over $\pm\Gamma_\phi$ region around $\phi$ peak  [use
Eq.~(\ref{braver}) with replacement $\omega\to\phi$]. The
evaluation of the excitation curve of the decays $\phi\to5\pi$ in
$e^+e^-$ annihilation performed according to Eq.~(\ref{sigomega})
(with the replacement $\omega\to\phi$) is plotted in
Fig.~\ref{fig13}. Notice that the ratio of the branching ratios of
two isotopic modes at $\phi$ peak is
\begin{equation}
{B_{\phi\to2\pi^+2\pi^-\pi^0}(m^2_\phi)\over
B_{\phi\to\pi^+\pi^-3\pi^0}(m^2_\phi)}=2.9,\label{phirat}\end{equation}
to be compared to the figure of 1.3 obtained from the simple
evaluation of the ratio of nonrelativistic phase space, see
Eq.~(\ref{omegarat}) with the replacement $m_\omega\to m_\phi$. In
the present case, the difference with the exact evaluation is
sizeable, because now the phase space model is inadequate due to
the strong $\rho$ and $\omega$ ($\rho\to\omega\pi\to4\pi$) resonance production in the
intermediate states. The $\phi\to5\pi$ excitation curve is plotted
in Fig.~\ref{fig13}.

In this respect, it is interesting to look at the dynamical
behavior of the specific  contributions to the $\phi\to5\pi$ decay
amplitudes in another way. Let us evaluate, for this purpose, the
contribution to $B_{\phi\to\pi^+\pi^-3\pi^0}$  of the diagrams
Fig.~\ref{fig1}, at the $\phi$
mass. (Notice that now $\omega$ in initial state  should be
replaced with $\phi$ in all the diagrams, and the effective
$g_{\phi\rho\pi}$ is understood at the corresponding expression,
while other couplings are related with it as is explained earlier
in this Section). The $\rho$ meson in these diagrams is resonant.
Indeed, choosing the averaged pion energy from the condition of
equilibrium as $\langle E_\pi\rangle=m_\phi/5$, one finds
that the invariant mass of four pions emitted in the transition
$\rho\to4\pi$ is $m_{4\pi}\simeq m_\rho$.
The evaluation gives  $B^{\rm resonant}
_{\phi\to\pi^+\pi^-3\pi^0}=2.1\times10^{-7}$. All the remaining
contributions with the non-resonant intermediate $\rho$ meson, see
Fig.~\ref{fig2}-\ref{fig4}, amount to $B^{\rm
non-resonant}_{\phi\to\pi^+\pi^-3\pi^0}=0.34\times10^{-7}$, which
constitutes $16\%$ of the resonant contribution. Notice
that the seemingly resonant diagrams Fig.~\ref{fig2} and
\ref{fig7} do not, in fact, possess this property, because three
pions produced from the transition $\pi\to3\pi$, push $\rho$ meson
away from the resonance. Indeed, the invariant mass of the pion pair
in the transition $\rho\to2\pi$ evaluated assuming the same average pion energy
as above,  falls into the interval $2m_\pi\leq m_{2\pi}
\leq0.41$ GeV, which is far from the resonance value.
The phase space averaged relative phase
between the resonant and non-resonant contributions calculated
with the help of given branching ratios and that given in
Eq.~(\ref{brphi}) is about $\delta=91^\circ$. Correspondingly,
similar calculations for another isotopic state
$2\pi^+2\pi^-\pi^0$ state give $B^{\rm resonant}
_{\phi\to2\pi^+2\pi^-\pi^0}=6.2\times10^{-7}$ from
Fig.~\ref{fig6}, $B^{\rm
non-resonant}_{\phi\to2\pi^+2\pi^-\pi^0}=0.70\times10^{-7}$ from
Fig.~\ref{fig7}-\ref{fig9}, and $\delta=89^\circ$. In the present case, the
non-resonant contribution constitutes about $11\%$ of the resonant one.
The above estimates  illustrate clearly  the dominance of
the diagrams with the resonant $\rho$
meson in the intermediate state in the decay $\phi\to5\pi$, because
the resonant and the smaller non-resonant contributions add
incoherently in the case of the $\phi\to5\pi$ decay. For
comparison, opposite situation takes place in the case of the
$\omega\to5\pi$ decay amplitudes, see the corresponding
calculations in Sec.~\ref{omegaresults}, where the smaller
non-resonant  contribution to the decay amplitude adds almost in
phase with the resonant one and by this reason is essential.

The relaxing the constraint Eq.~(\ref{betcond2}) to
$-1\leq(\beta_1+\beta_2-\beta_3)/4\beta_3\leq1$ analogous to that
discussed in the $\omega$ case implies even smaller deviations
of $B_{\phi\to5\pi}$ in comparison with the $\omega$ case, because
the terms in the amplitude which are sensitive to the parameter in the above
inequality, are almost incoherent with the dominant ones.
On the other hand, relaxing the constraint of the absence of the
point like $\phi\to\pi^+\pi^-\pi^0$ amplitude, see Eq.~(\ref{betcond1}), gives the following.
Using the KLOE data Ref.~\cite{kloe}, one can estimate the
combination characterizing the pointlike $\phi\to\pi^+\pi^-\pi^0$
vertex as $|3(\beta_1-\beta_2-\beta_3)/2\beta_3m^2_\rho|\simeq1$.
The evaluation of $B_{\phi\to5\pi}$, keeping the constraint
Eq.~(\ref{betcond2}), gives the results deviating
by $\pm8\%$ (depending on the sign of the above combination) from those obtained
under the constraint Eq.~(\ref{betcond1}).

All the above discussion shows that  the branching ratios of the decays
$\phi\to\pi^+\pi^-3\pi^0$ and $\phi\to2\pi^+2\pi^-\pi^0$
are determined within the conservatively estimated  accuracy 20\%
by the well studied OZI rule violating transition of $\phi$ meson to
the $\rho\pi$ state followed by the transition $\rho\to4\pi$  in the model independent way.

\section{Discussion and conclusion.}
\label{conclu}

In view of the fact that there are three (or even four, if one
includes radiative decays, see Ref.~\cite{hls1,hls2}) independent
constants in the effective chiral Lagrangian describing anomalous
decays of $\omega$ (and $\phi$) mesons, one can only consider some
scenarios of what may happen. We restrict ourselves by considering
only the strong decays. In principle, the study of the Dalitz plot
in the $\omega\to\pi^+\pi^-\pi^-$ decay allows to extract $c_3$
and $(c_1-c_2)/c_3$ by isolating the $\rho$ pole and non-$\rho$
pole contributions, because the density on this plot is
proportional, omitting the $\omega\rho$ interference term in the
$\pi^+\pi^-$ mass spectrum, to the factor
\begin{equation}
{d^2N\over dm_+dm_-}\propto\left|{1\over D_\rho(q_1+q_2)}+{1\over
D_\rho(q_1+q_3)}+{1\over D_\rho(q_2+q_3)}+3{c_1-c_2-c_3\over
2c_3m^2_\rho}\right|^2,\label{dalitz}\end{equation} where
$m_+^2=(q_1+q_3)^2$, $m_-^2=(q_2+q_3)^2$. Notice in this respect
that the combination of parameters of the low energy effective
Lagrangian entering in the non-$\rho$-pole term in
Eq.~(\ref{dalitz}) should be treated the low energy limit of all
possible contributions from the transitions
$\omega\to\rho^\prime\pi$, $\rho^{\prime\prime}\pi$ etc.
 If one assumes the direct transitions are
responsible for the decays of $\phi$ meson to the states
containing no strange quarks, the same will be true for the
parameters $\beta_{1,2,3}$ characterizing the OZI rule violating
decays $\phi\to3\pi$ and $\phi\to5\pi$. In the model of
$\phi\omega$ mixing, the $\phi\to5\pi$ decay amplitude contain no
additional free parameters as compared to the case of the
$\omega\to5\pi$ decay. It should be recalled that both models can
be, in principle, discriminated by the careful study of the
$\phi\omega$ interference minimum in the energy dependence of the
$e^+e^-\to\pi^+\pi^-\pi^0$ reaction cross section or by the ratio
of the leptonic widths of $\omega$ and $\phi$ mesons
\cite{ach95,ach93,snd03}. On the other hand, within the accuracy
of 20\%, the branching ratios of the $\phi\to5\pi$ decays can be
evaluated in a model independent way, see the discussion at the
end of Sec.~\ref{phiresults}.

The  excitation curves of the  decays  $\omega\to5\pi$ and
$\phi\to5\pi$  in $e^+e^-$ annihilation can be used to evaluate
the expected number of these decays at $\omega$ and $\phi$ peaks.
With the luminosity $L=10^{32}$ cm$^{-2}$ s$^{-1}$ at $\omega$
peak, one may hope to observe three events of the decays
$\omega\to\pi^+\pi^-3\pi^0$ and $2\pi^+2\pi^-\pi^0$ per each mode
bimonthly. With the same luminosity at the $\phi$ peak, the
observation of, respectively, 750 (250) $\phi\to2\pi^+2\pi^-\pi^0$
($\phi\to\pi^+\pi^-3\pi^0$) decays per month is feasible. Note
that the existing upper limit is
$B_{\phi\to2\pi^+2\pi^-\pi^0}<4.6\times10^{-6}$ ($90\%$ C.L.)
\cite{cmd200}. With the luminosity $L=500\mbox{ pb}^{-1}$ already
attained at $\phi$ factory DA$\Phi$NE \cite{kloe1}, one could gain
about 1685 events of the decay $\phi\to5\pi$ proceeding via chiral
mechanisms considered in the present paper. The possible
non-chiral-model background from the dominant decay $\phi\to
K_LK_S$, $K_L\to3\pi$, $K_S\to2\pi$ is well cut from the
considered chiral mechanism by macroscopic distances kaons fly
away. Rare decay $\phi\to\eta\pi^+\pi^-$ whose branching ratio was
estimated \cite{ach84,ach92} at the level
$B_{\phi\to\eta\pi^+\pi^-}\sim3\times10^{-7}$, is cut by removing
events in the vicinity of the $\eta$ peak in the three pion
distribution observed in the five pion events \cite{cmd200}.

In the present work, we neglect the contribution of the
$a_1(1260)$ meson. This is justifiable because both the
$\omega(782)$ and $\phi(1020)$ peaks are deep under the  threshold
of $a_1\pi$ production. As is known, the approach to chiral
dynamics based on HLS, allows to take the axial vector mesons into
account \cite{hls1,hls2}. This is the theme of future work.

\begin{acknowledgements}
The present study was partially supported by the grant
RFFI-02-02-16061 from Russian Foundation for Basic Research.
\end{acknowledgements}

\appendix*
\section{Relations expressing Lorentz scalar products through the Kumar variables}
\label{app}

In this Appendix, the relations expressing the Lorentz scalar
products $(q_i,q_j)$ through Lorentz-invariant variables are
presented. Given the pion momentum assignment according to
\begin{equation}\omega_q\to\pi_{q_1}\pi_{q_2}\pi_{q_3}\pi_{q_4}\pi_{q_5},\label{a1}\end{equation}
the eight Kumar variables \cite{kumar} are defined as
\begin{eqnarray} s_1&=&(q-q_1)^2,\nonumber\\
s_2&=&(q-q_1-q_2)^2,\nonumber\\
s_3&=&(q-q_1-q_2-q_3)^2,\nonumber\\u_1&=&(q-q_2)^2,\nonumber\\
u_2&=&(q-q_3)^2,\nonumber\\u_3&=&(q-q_4)^2\nonumber\\t_2&=&(q-q_2-q_3)^2,\nonumber\\
t_3&=&(q-q_2-q_3-q_4)^2.\label{kumvar}\end{eqnarray} Associated
with them, but not independent, are the following:
\begin{eqnarray}
s^\prime_2&=&(q_1+q_2)^2,\nonumber\\
s^\prime_3&=&(q_1+q_2+q_3)^2,\nonumber\\
s^\prime_4&=&(q-q_5)^2,\nonumber\\
t^\prime_2&=&(q_2+q_3)^2,\nonumber\\
t^\prime_3&=&(q_2+q_3+q_4)^2.\label{kumvarpr}\end{eqnarray}Then
the greater part of the Lorentz scalar products of the pion
momenta can be  expressed through the variables Eq.~(\ref{kumvar})
and (\ref{kumvarpr}):
\begin{eqnarray}
(q_1,q_2)&=&{1\over2}(s^\prime_2-m^2_1-m^2_2),\nonumber\\
(q_1,q_3)&=&{1\over2}(s^\prime_3-s^\prime_2-t^\prime_2+m^2_2),\nonumber\\
(q_1,q_4)&=&{1\over2}(t_2-t_3-s_3+m^2_5),\nonumber\\
(q_1,q_5)&=&{1\over2}(t_3-m^2_1-m^2_5),\nonumber\\
(q_2,q_3)&=&{1\over2}(t^\prime_2-m^2_2-m^2_3),\nonumber\\
(q_4,q_5)&=&{1\over2}(s_3-m^2_4-m^2_5).\label{qijsimp}\end{eqnarray}
The remaining scalar products
\begin{eqnarray}
(q_3,q_5)&=&{1\over2}(s_2-s_3-m^2_3)-(q_3,q_4),\nonumber\\
(q_2,q_4)&=&{1\over2}(t^\prime_3-t^\prime_2-m^2_4)-(q_3,q_4)\label{qijcomp}\end{eqnarray}
can be expressed through $(q_3,q_4)$. The latter, using the method
of invariant integration outlined in Appendix D of
Ref.~\cite{kumar}, can be found as
\begin{equation}
(q_3,q_4)={1\over2}\left[\alpha(s-u_2+m^2_3)+\beta(u_1-t_2-m^2_3)+\gamma(s_2-s_3-m^2_3)\right],
\label{q23}\end{equation}where
\begin{eqnarray}
\alpha&=&{1\over\Delta_M}(Ft_2s_3+BCG+ACH-t_2BH-C^2F-As_3G),\nonumber\\
\beta&=&{1\over\Delta_M}(ss_3G+ABH+BCF-B^2G-sCH-As_3F),\nonumber\\
\gamma&=&{1\over\Delta_M}(st_2H+ABG+ACF-t_2BF-sCG-A^2H),
\label{albetgam}\end{eqnarray}and
\begin{eqnarray}
A&=&{1\over2}(s+t_2-t^\prime_2),\nonumber\\
B&=&{1\over2}(s+s_3-s^\prime_3),\nonumber\\
C&=&{1\over2}(s_3+t_2-m^2_1),\nonumber\\F&=&{1\over2}(s-u_3+m^2_4),\nonumber\\
G&=&{1\over2}(t_2-t_3+m^2_4),\nonumber\\
H&=&{1\over2}(s_3+m^2_4-m^2_5),\nonumber\\
\Delta_M&=&st_2s_3+2ABC-B^2t_2-C^2s-A^2s_3.\label{abcde}\end{eqnarray}
In the above formulas, $m_i$, $i=1,...5$, are the masses of final
pions.

\begin{figure}
\includegraphics{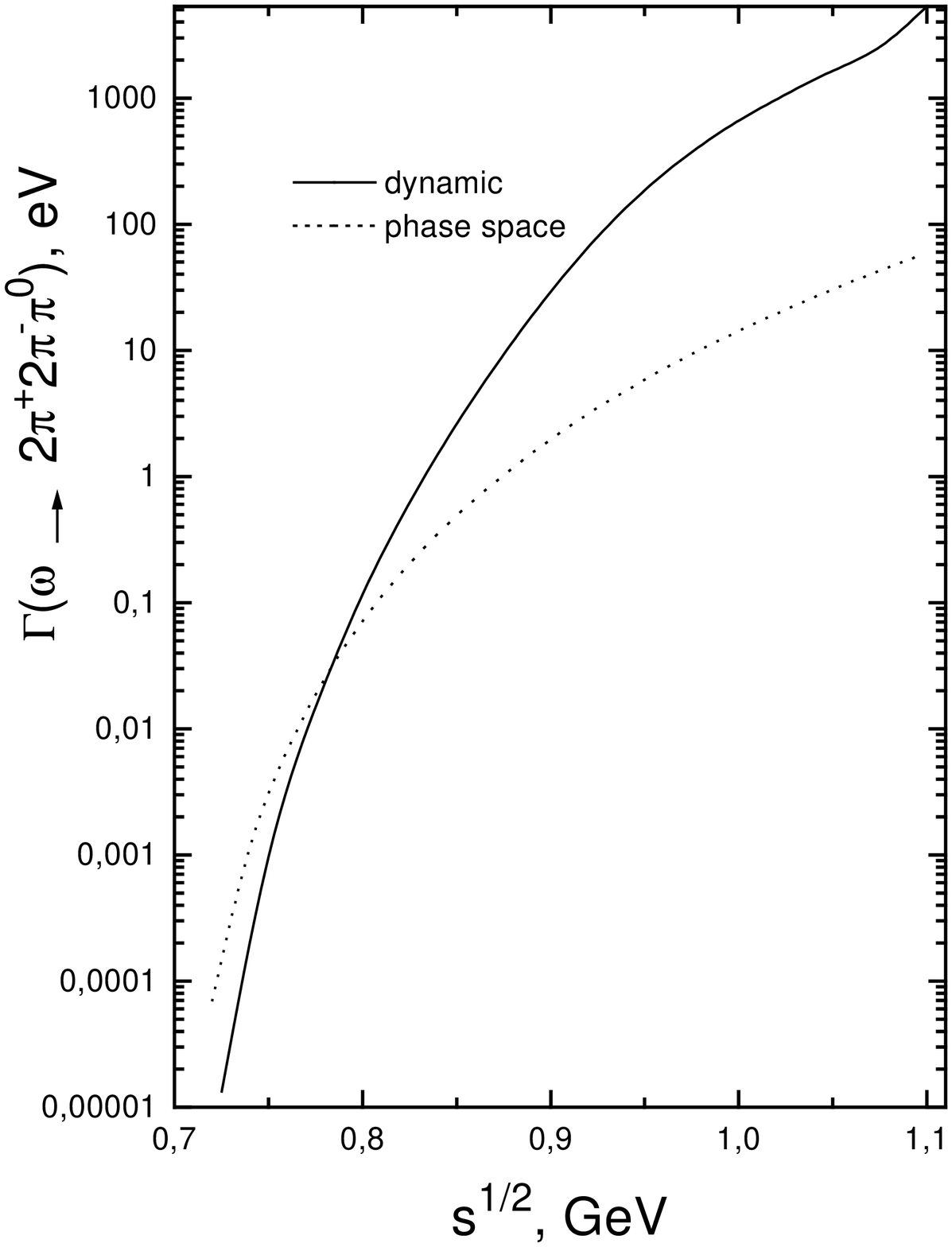}
\caption{\label{fig12}The energy dependence of the $\omega\to2\pi^+2\pi^-\pi^0$
partial width.}
\end{figure}
\begin{figure}
\includegraphics{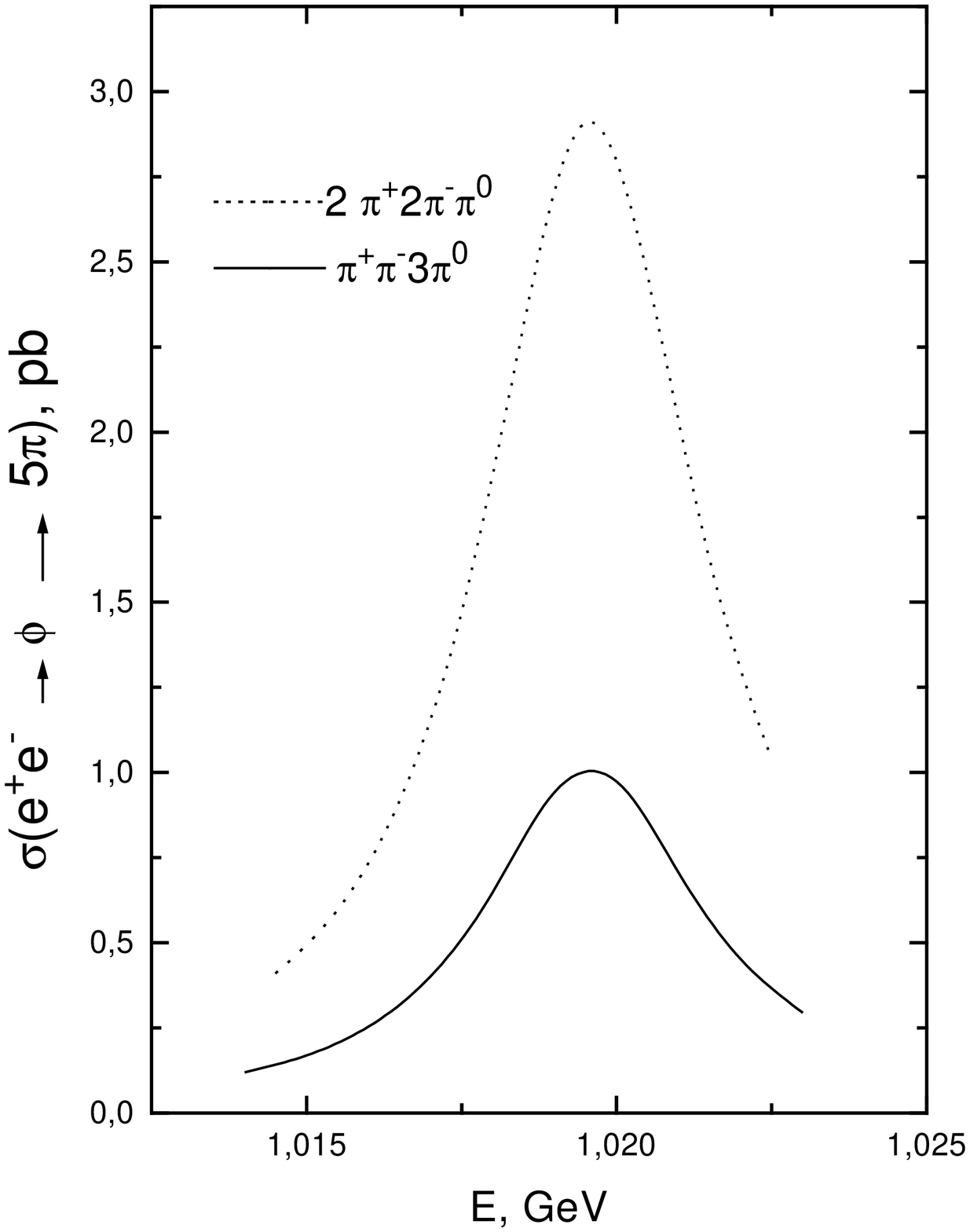}
\caption{\label{fig13}The excitation curve of the decays  $\phi\to5\pi$ in $e^+e^-$
annihilation.}
\end{figure}
\end{document}